*Alexander D. Smirnov*

**To Buy or Not to Buy, That's Not the Question:**

**A Simple Model of Credit Expansion**

**Smirnov, A.D.,** *To Buy or Not to Buy, That's Not the Question: A Simple Model of Credit Expansion.* Preprint WP2/2012/0…. – M: NRU HSE, 2012. – 42 p.

The proposed model is aimed to reveal important patterns in the behavior of a simplified financial system. The patterns could be detected as regular cycles consisting of debt bubbles and crises. Financial cycles have a well defined structure and form periodic sequences along the axis of credit expansion while retaining stochastic nature in terms of time. Bubbles are defined as "large asset price deviations" from their fundamental value, and crises, if happen, would represent huge losses of financial wealth. Regular sequences of bubbles and crises in the model are explained via behavior of market participants whose collective actions facilitate either emergence or postponement of crises due to events as investors' self-imposed restrictions upon debt accumulation. The paraphrase of famous Hamlet's dilemma exposes the illusory character of investors' attempt to avoid crises: financial catastrophes, even being postponed, are inevitable under the regime of credit expansion. It was shown, in particular, that the probability of default grows in coherence with the increasing money issuance, and at the point of a total collapse it reaches the unit value. The model clearly discerns phases of normal investing, speculation and a Ponzi game which have the same meaning as in the "financial instability hypothesis" elaborated by H. Minsky. The model while supporting the assessment of crises inevitability under regime of a credit expansion, does not share the Austrian School doomsday predictions regarding the exact dates of such events.

**Smirnov, Alexander D.,** Professor, National Research University – Higher School of Economics, Pokrovsky Boulevard 11, Moscow 101987, Russia, tel. 772 9590*2175, e-mail: adsmir@hse.ru; alexandersmirnov16@gmail.com

"There is no means of avoiding the final collapse of a boom brought about by credit expansion. The alternative is only whether the crisis should come sooner as a result of the voluntary abandonment of further credit expansion, or later as a final and total catastrophe of the currency system involved."

**Ludwig von Mises**, *Human Action. A Treatise on Economics*

# Contents

**Methodological remarks on crises modeling**

Global financial crisis of 2007-2009 had brought about huge losses imposed upon financial institutions and non-financial corporations, producers and consumers. Customers were overloaded with houses, once abundant credit evaporated, world trade and production were severely disrupted, and unemployment dramatically increased[1]. Though, as usual, "this time was different", the real market failures were preceded by a credit crunch as it had always been in economic history.

An important aspect of the last crisis, comprising its devastating effects upon the intellectual projection of national economies and finance, is to be stressed specially. In this respect, one of the consequences of the credit crunch was a spectacular demise of the concept of "market fundamentalism" (Soros, 2006).It included, naturally, its macroeconomic and financial projection – the "representative agent" model. Barrage of criticisms and negations addressing the latter were justified by a remarkable failure of theoretical attempts to detect and foresee the upcoming crisis (Economist, 2009). It should be noted that such attempts were not, mildly speaking, encouraged, especially within the then dominant theory of "rational expectations". Even more, the mere subject of crises was effectively excluded from the main body of that concept (Lucas, 2003). On the other hand, being calmed down by a prolonged period of "Great Moderation", economists of the mainstream school had lost, to some degree, their interests in the analysis of cycles and extreme (or "fat tail") events like systemic bubbles and crises. Quite understandably, once critical phenomena had been declared as virtually nonexistent, the "representative agent" model, by implication, was elevated to the class of universality. Hence, the entire, the then dominant, theoretical paradigm was doomed to ignore the complexity of a modern financial system and irregularities of extreme events being born by its evolution, especially on a macrolevel. The mere inability of the mainstream theory to explain and predict the credit crunch of 2007-09 sparked the interest of practitioners to the apocryphal ideas in economics, the Minsky instability hypothesis in particular (Yellen, 2009).

In our opinion, the inadequacy of a "rational investor" model was rooted in its unreserved reliance on a methodology of reductionism which, in our view, could be ascribed as the *causa sine* of the impotence of the model, its failure to identify and predict, in advance, credit crunch of 2007. As known, the reductionism declares, unconditionally, the existence of similarity of any system to its (typical) element, and in a guise of a "rational" investor, that doctrine became a dominant theoretical engine in a science of finance. The "rational investor" model dominance has been continued for several decades, in spite of the fact that its inadequacy to the actual financial markets was well known and extensively documented (Eichengreen, 2003; Mandelbrot, 2005; Moessner, 2010).

Unconditional similarity contradicts to the modern complex systems theory, especially to its parts which are focused on the analysis of the so called "critical points" where a general system transforms its quality and behavior (Stanley et al, 2003). The methodology of reductionism turned out to be largely at odds with financial activity, for, as a rule, financial markets appear to be scale free only under very special conditions, near some critical point, for example. Hence the assertion of a system's similarity to its element, while being relevant under some conditions, cannot be justified in general. A huge body of evidence suggests that market transformations, similar to credit crunches, take place typically at critical points. Since financial system evolves largely as a laminar flow while infrequently, under some specific conditions, producing bursts of turbulence, the endogenous mechanism of such "switches" is of great importance. Possibly, this mechanism could be viewed as a transformation of heterogeneous ("normal") markets into homogeneous ones as a result, for example, of a bubble burst due to sudden disappearance of buyers that is trailed by market illiquidity.

---

[1] The author is grateful to the participants of seminars at the *London School of Economics*, the *National Research University – Higher School of Economics (Moscow)* and some financial firms in the City. He is especially thankful to Professors *Avinash Dixit*, *Charles Goodhard, Fuad Aleskerov* and *Emil Yershov* whose comments were important for the model elaboration. Yet nobody, except the author, bears any responsibility for possible inconsistencies of the paper.

The rational agent model, as it became increasingly noticeable, has proved to be incapable to identify properly and filter out extreme events that appear near critical points, including bubbles or crises. For example, Gaussian distribution of events which is an imminent part of the "rational investor" model under uncertainty, returns a practically zero probability of a would be crisis. This result, scientifically quite correct, served, in its turn, as a solid argument supporting the virtual nonexistence of extreme events in finance[2]. Logically, though, it is a false inference since empirical data should be correctly identified as the evidence of "no crises". Hence the error might appear as a product of logical fallacy – empirical data containing "no evidence of crises" were interpreted, wrongly, as containing an "evidence of no crises" (B. Russell, Chicken Paradox, 1924)[3]. The latter, quite naturally, implied the conclusion of the "crises nonexistence", at least in the modern finance. Looking from this angle, an assessment of a rational investor model as a universal theory resembles the error of the second kind in testing of statistical hypotheses: instead of rejecting the wrong null hypothesis of "nonexistence of crises", the latter was put into foundation of a theory that pretended to be a universal one.

Devastating failure of the rational investor model being applied to the study of credit crunch 2008-09 has validated an unprecedented search for the "New Economic Paradigm" (Stiglitz, 2010) including comprehensive revision of basic premises of a financial science[4]. This search was initiated and followed by leading scientists including J. Stiglitz, P. Krugman, G. Akerlof and R. Schiller, D. Farmer - to mention just a few. An important avenue of these intellectual activities is an investigation of the mechanism of financial bubbles and crises, their study as complex, uncertain and hierarchically organized phenomena. Although bubbles and crises are macro-financial phenomena, financial activity on a macro- level, contrary to microfinance, has been, for different reasons, a relatively underdeveloped part of modern finance.

As one of the participants of the 2011 INET Conference remarked, the " new economic thinking means reading of the old books". It is a half-jest, in fact, for the principles of a new economic thinking were laid firmly down in works of J. M. Keynes, I. Fischer, H. Minsky, H. Simon, B. Mandelbrot. Their ideas, as well as scientific results of some contemporary researchers, especially those who emphasized the importance of studying phenomena of "animal spirit", "herding" or "irrational exuberance" has formed, in effect, foundations of a macrofinancial theory, including the analysis of bubbles and crises. In our view, the new paradigm should go along the avenues of a complex system methodology. The latter provided researcher with a wide spectrum of methods and models (*Encyclopedia*, 2009) that were gratefully acknowledged, for often the same approach have long been used the economic theory itself. The intertwined fields of research on copula models, firms growth, econophysics, portfolio defaults, currency crises and sovereign defaults could be mentioned in this context (Vasicek, 1987; Lux, 2006; Li,2000; Hull, 2011).

Theoretical generalizations of a financial history, as it had been investigated and described by such prominent minds as L. vonMises, F. vonHyek, M. Friedman and A. Schwartz, J.K. Galbraith, C. Kindleberger,

---

[2] On the "Black Monday", October 16, 1987, Standard&Poor's500 Index had dropped about 34 standard deviations which for Gaussian distribution corresponds to the probability of $2.6 \times 10^{-256}$. In practice, such an extreme event could had never happened (Stanley,et al,2003).
[3] A persuasive exposition of the "Chicken Paradox" can be found in (Taleb, 2010).
[4] There is some irony in the name of "the modern finance" which the mainstream financial theory took possession of to identify itself. In fact, its foundations had been formulated about at least half-a-century ago. That does not mean, of course, that all the hypotheses underlying the mainstream theory are wrong but, on the other hand, the mere proclaiming their adequacy to the complexities of modern financial markets looks a bit arbitrary, and imposing some inherently dangerous consequences.

P.Krugman, B. Eichengreen - to mention just a few - might be viewed as a reconstruction of a single realization of a stochastic process. The latter could possess many features with some of them totally unknown. The development of a financial system, being a flow of funds in time, cannot, generally, be considered a laminar process. In the long run it had always been subject to sudden and aperiodic bursts of heavy turbulence, caused by internal and external shocks. Episodes of the turbulence, though being relatively rare (the Poisson process), have been subject of thorough investigation since long ago economists had been aware of their strong impact upon the system. If amplitudes of fluctuations were extremely large, such events would be called crises. The importance of their study was emphasized by numerous facts of devastating influence of crises upon the structure and behavior of a financial system due to huge losses imposed upon the wealth of nations, both private and public.

Financial bubbles were always forerunners of crises. The modern financial system analysis takes the view on bubbles in a context of "large asset price deviations" from their fundamental value (Rajan, 2005). Similar ideas were developed and investigated in (Turner, 2010) who wrote that "all liquid financial markets are susceptible to unstable divergence from equilibrium values". Such a general system approach takes essentially into account market interactions responsible for qualitative changes in the aggregate system. Irrationality or herding of financial investors bears responsibility for asset price divergence that under particular conditions emerges and brings about a system's collapse. The possibility of persistent deviations of asset prices from their fundamental value was shown in for AR (1) stochastic processes (Campbell and Shiller, 1992). This paper makes an attempt to describe prices divergence at the critical point (where a system becomes singular) by appealing to investors' actions and motivations on the macrofinancial scale. Investigation of a system's behavior around the point of singularity is of vital importance since it would provide essential clues to our understanding of "how markets fail" (Cassidy, 2009).

The complex system approach in finance could be described through the concept of entanglement. The concept of entanglement bears the same features as a definition of a complex system given by a group of physicists working in a field of finance (Stanley et al, 2003). As they defined it – in a complex system all depends upon everything. Just as in the complex system the notion of entanglement is a statement acknowledging interdependence of all the counterparties in financial markets including financial and non-financial corporations, the government and the central bank. How to identify entanglement empirically? Stanley H.E. *et al* formulated the process of scientific study in finance as a search for patterns. Such a search, going on under the auspice of "econophysics", could exemplify a thorough analysis of a complex and unstructured assemblage of actual data being finalized in the discovery and experimental validation of an appropriate pattern. On the other side of a spectrum, some patterns underlying the actual processes might be discovered due to synthesizing a vast amount of historical and anecdotal information by applying appropriate reasoning and logical deliberations. The Austrian School of Economic Thought which, in its extreme form, rejects application of any formalized systems, or modeling of any kind, could be viewed as an example. A logical question follows out this comparison: Is there exists any intermediate way of searching for regular patters in finance and economics? It should be pointed out, that in our opinion, though these patterns imply the existence of some stable structures, the latter were not necessarily could be exploited, say, by practical investors[5].

Importantly, patterns could be discovered by developing rather simple models of money and debt interrelationships. Debt cycles were studied extensively by many schools of economic thought (Akerlof, Shiller, 2009). Some important characteristics of such a cycle could be revealed in the financial system

---

[5] Econophysical studies show that in many cases persistent features of random processes of non-Gaussian kind do not support arbitrage profits (Lillo, Farmer, 2004)

development at the turn of the century. The modern financial system worked by spreading risk, promoting economic efficiency and providing cheap capital. It had been formed during these years as bull markets in shares and bonds originated in the early 1990s. These markets were propelled by abundance of money, falling interest rates and new information technology. Financial markets, by combining debt and derivatives, could originate and distribute huge quantities of risky structurized products and sell them to different investors. Meanwhile, financial sector debt, only a tenth of the size of non-financial-sector debt in 1980, became half as big by the beginning of the credit crunch in 2007. As liquidity grew, banks could buy more assets, borrow more against them, and enjoy their value rose. By 2007 financial services were making 40% of America's corporate profits while employing only 5% of its private sector workers. Thanks to cheap money, banks could have taken on more debt and, by designing complex structurized products, they were able to make their investment more profitable and risky. Securitization facilitating the emergence of the "shadow banking" system foments, simultaneously, bubbles on different segments of a global financial market.

Yet over the past decade this system, or a big part of it, began to lose touch with its ultimate purpose: to reallocate deficit resources in accordance with the social priorities (BCBS, 2004). Instead of writing, managing and trading claims on future cashflows for the rest of the economy, finance became increasingly a game for fees and speculation. Due to disastrously lax regulation, investment banks did not lay aside enough capital in case something went wrong, and, as the crisis began in the middle of 2007, credit markets started to freeze up (Inquiry Report, 2011). Qualitatively, after the spectacular Lehman Brothers disaster in September 2008, laminar flows of financial activity came to an end. Banks began to suffer losses on their holdings of toxic securities and were reluctant to lend to one another that led to shortages of funding system. This only intensified in late 2007 when Nothern Rock, a British mortgage lender, experienced a bank run that started in the money markets. All of a sudden, liquidity became in a short supply, debt was unwound, and investors were forced to sell and write down the assets. For several years, up to now, the market counterparties no longer trust each other. As Walter Bagehot, an authority on bank runs, once wrote:" Every banker knows that if he has to prove that he is worth of credit, however good may be his arguments, in fact his credit is gone." (The Economist, 2008). In an entangled financial system, his axiom should be stretched out to the whole market. And it means, precisely, financial meltdown or the crisis.

The most fascinating feature of the post-crisis era on financial markets was the continuation of a ubiquitous liquidity expansion. To fight the market squeeze, all the major central banks have greatly expanded their balance sheets. The latter rose, roughly, from about 10 percent to 25-30 percent of GDP for the appropriate economies. For several years after the credit crunch 2007-09, central banks bought trillions of dollars of toxic and government debts thus increasing, without any precedent in modern history, money issuance. Paradoxically, this enormous credit expansion, though accelerating for several years, has been accompanied by a stagnating and depressed real economy . Yet, until now, central bankers are worried with downside risks and threats of price deflation, mainly. Yet.

**The model overview**

This preprint is a completely revised version of the previous paper (Smirnov, 2011). Methodologically the proposed model is aimed to reveal important patterns in the behavior of a simplified financial system viewed as an intertwined feedback between money and debt collaterized by real resources. Such patterns are detected as cycles consisting of debt bubbles and crises. Financial cycles have a well defined structure and form periodic sequences along the credit expansion though in terms of time they are of stochastic nature. Bubbles are defined as "large asset price deviations" from their fundamental value while crises represent huge losses of financial wealth. Regular sequences of bubbles and crises are explained in the model via behavior of market participants whose collective actions facilitate either crises or their postponement due to self-imposed restrictions upon the debt accumulation. The Hamlet dilemma paraphrase exposes the illusory character of investors' attempts to avoid crises: even being postponed financial catastrophes are inevitable under regime of credit expansion. It was shown that probabilities of default are growing along the increasing money issuance reaching the unit value in the case of a total collapse. The model distinguishes phases of normal investing, speculation and a Ponzi game as in the "financial instability hypothesis" elaborated by H. Minsky (Minsky, 2008). An important reservation should be made concerning the model validity. Since it embraces the debt markets only, the inflationary impact of the credit expansion is, in effect, reflected in the growing value of debts. The term "inflation", in a particular sense, could be conceived as a synonym to the phenomenon of a credit expansion which is often the case in the works of the Austrian scholars (vonMises, 1996). It does not in contradiction, though, with its total absence on the real markets where inflation is registered as changes in prices of goods and services. If the economy in a liquidity trap, then there could be no inflation at all on the goods markets while asset prices are skyrocketing. Hence, supporting the assessment of the crises inevitability under the credit expansion, the model does not share the Austrian School doomsday predictions of exact dates of such events.

The model captures at least four stylized facts of the modern financial system performance, shortly described above. First, the model reflects dramatic changes in proportions between the real and financial markets, namely the fact, that the modern financial system becomes many times larger than the economy *per se*. Second, it describes financial evolution as a process transforming simple bank intermediation into "alternative" banking. Third, all the money issued, in the model are spent on the new debt acquisition like in the process of quantitative easing, QE, where new money are issued due to debt purchases of the central bank. Four, in the process of money issuance the model captures its feature of the excess money accumulation. These viable features of the actual evolution of a financial system are important in understanding the financial cycle pattern as it was thoroughly investigated both within the mainstream economic thought (J.M. Keynes, H. Minsky, C. Kindelberger, P. Krugman, B. Eichengreen), as well as by some prominent Austrian scholars (L. von Mises, F. Hayek, D. French).

The financial market transition from laminar to turbulent regimes reveals, in terms of money issuance, the regular cyclic pattern of subsequent credit crunches. The latter, in terms of time, appears to be a rather irregular one. Due to stochastic liquidity issuance, the model demonstrates (according to a biased random process) a financial system evolution towards the shadow banking system. This transition is an intermittent process for in its course bubbles emerge and burst sporadically, thus forming a cyclical process. The latter, on the macrofinancial scale, could be represented via large asset prices deviations. That theoretical concept reflects the switching of investors' preferences from the market to the expected debt value. The structural change in investors' orientation inherently leads to systematic asset price overvaluations. The magnitude of the asset price divergence reaches its maximum at the critical point of liquidity issuance being accompanied with a nonzero probability of systemic collapse. The latter could be associated with a relatively modest market correction. The system, if survived at the critical point, continues to evolve via credit expansion until the growing amount of liquidity erodes the purchasing power of money completely. The system comes to its ultimate and total collapse when the debt reaches its zero value which might be viewed as an outcome of financial investors herding that brings about the so called "fat tail" financial event. Some important features of these complex processes are reproduced in the proposed model.

Stochastic differential equations in the model were solved via methodology elaborated by A.Dixit and R.Pindyck (Dixit and Pindyck, 1994). Solutions made relevant variables to be simple power functions combinations of which reveal regular cyclic patterns in the system's behavior described in terms of money issuance. The same approach was helpful in studying the phenomenon of "large price deviations" considered as one of the major sources of the cyclical pattern. Processes of herding in the financial market were represented, formally, as solutions to equations of the "trivial" nonlinear programming problem. The model included two types of financial risks. By hedging out the risky component of money issuance, the system was converted into a deterministic one. Risks to default were treated along the lines of the "financial triangle" approach which is common in estimation of probabilities to default. In the last part of the paper, financial bubble was viewed as a percolation process of forming clusters among the debt buyers. Models of such type are seemed adequate in investigating of microfinancial interactions among investors near the critical point (Smirnov, 2010). Since models of that kind are constructed within a different methodological approach, they are beyond the scope of the paper.

The system behavior could be summarized as follows. Its initial configuration is given as a Keynesian two-component market consisted of money and debt that evolves as interactions among financial investors, central bank and the government. The debt market is "broad" and "deep" initially, with many heterogeneous buyers and sellers (Malkiel, 2012). Private investors' purchases and sales of assets influence and determine the debt market value. The government, on its part, sells public debt while providing, at some levels of money issuance, deposit insurance which is a part of the put-to-default guarantees of the system stability. The central bank performs monetary policy which is subject to random shocks of internal and external nature. Financial markets via gradual random increases of market liquidity evolve towards the "shadow banking system". gradual random increases of market liquidity . The latter term in the model describes the system where decreasing put-to-default guarantees are substituted for additional credit of the call option type. Thus the central bank policy is viewed as a stylized description of quantitative easing policy, channeling the huge influx of money into financial markets, while leaving the real ones relatively immune to the unprecedented monetary stimulus.

Under "normal" market conditions the system dynamics goes on as a standard debt monetization process. The latter was modeled via the debt value equation reflecting interactions between money and debt. Debt value in aggregate is considered as a function of a random liquidity issuance, the latter being subject to the lognormal distribution (geometric Brownian motion). Monetary policy is modeled as a simple stochastic differential equation with a positive small drift parameter. Random process of liquidity issuance by the central bank brings about changes to the value of financial claims; hence all the model variables depend upon money issuance except the par value which is assumed to be constant. The credit availability facilitates additional debt purchases changing via call option written on the expected debt value. The face value of a debt is viewed as a riskless debt. The total debt guarantees, in their turn, were viewed as a put option also being written on its expected value. Call and put options are imbedded into debt making its market value a structurized financial product. It has a dual representation: either as a difference between its expected value and the call option, or equivalently, as the par value minus the put-to-default option. The above said statement is the macrofinancial interpretation of a well known put - call equivalence theorem. A modified representation of that equivalence (using the basic accounting equation) allows for performing asset and equity value estimations. Value of equity in the system appears to be equal to the sum of put and call options.

By implying persistent substitution of the market debt value for its expected value, the model characterizes an emergence of a financial bubble. The author believes that its origins are rooted in the normal market conditions as was noted by (Cooper, 2008). The model performance is facilitated by a persistent drift in liquidity issuance that brings about changes in all financial variables except the par debt value. The latter is a constant viewed as a riskless quantity being supported by the real resources. Financial system might arrive at the critical point under two different regimes. By definition, normal regime of financial market implies the absence of investors' herding in the debt purchases. Thus this regime allows for capital owned by investors in amounts returning probability of a crisis being strictly less than unity. The crisis under these conditions could be associated with the market correction.

Otherwise, a hectic financial activity that is going on along unbounded credit expansion could be transformed by herding into autocatalytic process that, if being subject to accumulation of a new debt, might drive the entire system at a total collapse. From a financial point of view, this systemic collapse appears to be a natural result of unbounded credit expansion which is 'supported' with the zero real resources. Since the wealth of investors, as a whole, becomes nothing but the 'fool's gold', financial process becomes a singular one, and the entire system collapses. In particular, three phases of investors' behavior - hedge finance, speculation, and the Ponzi game – investigated and defined by H. Minsky, could be easily identified in the model as a sequence of sub-cycles that unwound ultimately in the total collapse.

## Basic structure of a financial market

As stated above, financial markets as a whole operated initially under a "normal" regime. By this term we mean markets dominated by rational investors buying and selling bonds and money. In a simple deterministic context, the total value of financial assets, $A(t)$, could be defined as a simple Keynesian two-component structure. Hence, at any time, $t$, it is equal to the sum of money, $M(t)$, and the expected value of debt, $B(t)$:

(1)  $A(t) = M(t) + B(t).$

Each variable in equation (1) is assumed to be continuous and twice differentiable function of time. For any infinitesimally short period borrowers in aggregate are to service their debt at the market (risk-adjusted) rate of return, $\mu$, subject to equation $dA = \mu B \, dt$. In order to do that, all the creditors, in aggregate, are to receive their periodical (coupon) income, $m(t)dt$, and acquire new debt, $dB$. Since money does not earn revenue, the general creditor-borrower balance corresponds to the following differential equation:

(2)  $\mu B(t)dt = m(t)dt + dB.$

Given initial debt value, $B(0)$, and a constant parameter $\mu$, equation (2) can be easily solved as

(3)  $B(t) = B(0) \exp[\mu t] - \int_0^t m(u) \exp[-\mu(u - t)] \, du.$

The debt value, as given by (3), may increase unboundedly in time. On the other hand, if at some future date, $t = t^*$, debt is to be redeemed, $B(t^*) = 0$, its current value should equal to

(4)  $B(0) = \exp[\mu t] * \int_0^t m(u) \exp[-\mu u] \, du,$

which, for a continuous flow of constant coupon payments, is a representation of simple annuity:

(4')  $B(0) = \frac{m}{\mu}(1 - \exp[-\mu t]).$

In the following the expected debt value is considered to be twice differentiable function, $B(t, s_t) = B(s_t)$, of money issuance (money density at any moment of time, $s_t$). There are two aspects to be explained within such an assertion. On the one hand, the debt value dependence upon money issuance, $B(t, s_t) = B(s_t)$, by making the debt maturity profile irrelevant, greatly simplifies the model.

Secondly, from the economic point of view, the study of a debt value dependence upon money has a long scientific tradition that could be traced back at least centuries ago, especially, if being associated with the Austrian School of Thought. The latter established, rather firmly and persuasively, the causal links between money and debt by assembling a huge body of information, mainly of anecdotal character, about the excess money as precursors of the extreme economic events like debt overvaluation and financial bubbles. Historically, money excesses were always major cause to many extreme economic phenomena, ranging from the Tulipmania to the Great Depression. For example, a sudden rage for tulips in 17[th] century Holland, when a tulip bulb was worth more than a large mansion, is a well known fact. Though such a rage should had been supported with money, it was less known that the excessive issuance of florins in the period prior to the crisis had been the real factor that caused the tulipmania crash later (French, 2009). The credit crunch of 2007-09 could be analyzed along the same lines, for the excess liquidity, having been formed on global markets, gave rise to a subsequent credit meltdown.

Though not unanimously acknowledged, the "Austrian tradition" of money-debt studies seems to be very convenient in modeling of financial processes. *Pecunia pecuniam parere non potest* (money cannot beget money). In the same way, the money-debt relation cannot be treated in a straightforward manner, in a sense that money begets debt. Anyway, the contrary is untrue, as well. Rather, money (money issuance in the model) forms conditions imposed upon financial markets. Applying a physical analogy of interrelations among the volume, pressure and temperature, it could be said that money or liquidity plays the role of a temperature on the financial markets. Similar to increases in the volume due to higher temperature (given pressure), the larger liquidity issuance would increase financial activity measured, for example, by the volume of contracts, and, as it well known, the latter forms the precise meaning of the notion of liquid market.

Due to uncertainty prevalent on financial and real markets, it is necessary to treat debt dynamics as a random process. The factor of uncertainty in the model is taken into account in the simplest possible way. Namely, the total debt value is assumed to evolve according to the following stochastic equation:

(5) $\quad dB = [\mu B(s_t) - s_t]dt + \sigma B(s_t)dz_t,$

which is a biased geometric Brownian motion that is used as a standard tool in financial modeling.

**Stochastic money issuance**

The issuance of money (monetary base) is viewed in the model as an exclusive prerogative of a central bank which issues money stochastically due to uncertainty prevailing in the financial and real markets. Figure 1 shows the weighted index of market liquidity (Gieve, 2006) whose changes in time resemble a standard stochastic process. Very similar performance is demonstrated on Figure 2 which represents dynamics of the currency component of the U.S. money supply.

Money issuance is supposed to be a random process depending upon time $t$, $s = s_t$. Due to decomposition into deterministic and stochastic components, the rate of money issuance could be represented as a geometric Brownian motion:

(6) $\quad \frac{ds}{s_t} = a\,dt + \sigma dz_t$

where $a > 0$ is a positive drift and $\sigma > 0$ is a volatility parameter. Stochastic differential equation (6) can be solved along the standard procedure that gives rise to the following random process for money issuance:

(7) $\quad s_t = s_0 \exp[(a - 0.5\sigma^2)t + \sigma z_t]$

where the term $z_t = \int_0^t dz_u$ is the Ito integral of a random noise. Meanwhile, the nonrandom function

(8) $\quad \langle s_t \rangle = s_0 \exp[at]$

is used as a representation of the expected money issuance for a period $t$.

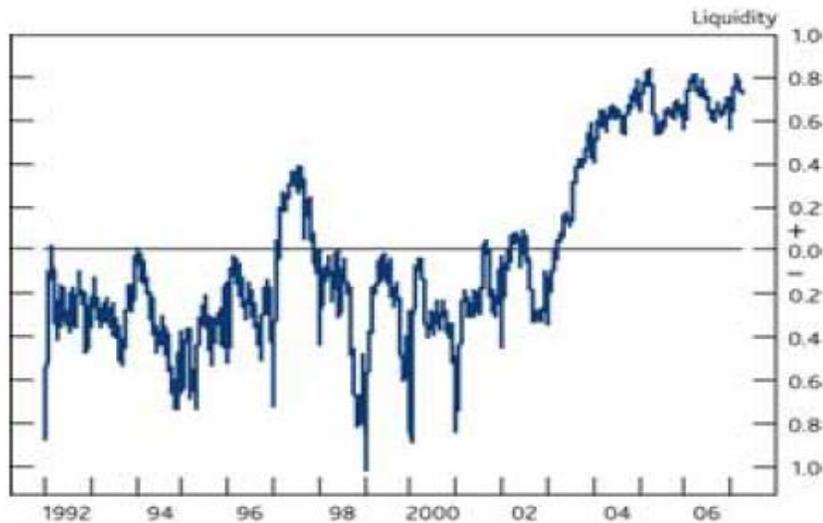

Figure 1. The weighted index of market liquidity measures for 1992-2006.

The monetary policy - quantitative easing, QE, for example - is performed by the central bank in accordance with (7) subject to a stochastic noise due to uncertainty prevalent on financial and real markets. Market participants expect money issuance in amounts given by nonrandom function (8). In accordance with the logic described above, financial investors, due to money issuance, acquire new debt persistently in a random fashion. In the micro-finance modeling money is usually considered to be a riskless asset which pay out zero rate of interest (Farmer, 2000). On the macrolevel money issuance is evidently a stochastic process. For example, the most popular "game of a town" among investors is the prediction of refinancing rates to be used in a subsequent period by relevant central banks. The conflict between these two approaches does not show up, since, firstly, private investors monitor averaged amounts of money and, second, they hedge risks out of their portfolios.

Hence the assumption of riskless money is realized indirectly: all the model variables appear to be intertwined deterministically though money issuance behaves as a random process.

Generally, volatilities of debt and liquidity processes in (5) and (6) are different but this technical detail is avoided for the moment. The debt value (5) and the liquidity dynamics (6) equations are augmented by standard connections among the major rates of return. By definition, the risk-adjusted rate of return, $\mu$, on financial (debt) assets is equal to the sum of the current yield, $\delta = \frac{m}{B}$, and the rate of capital appreciation (loss), $a = \frac{dB}{B}$:

(9) $\mu = \delta + a.$

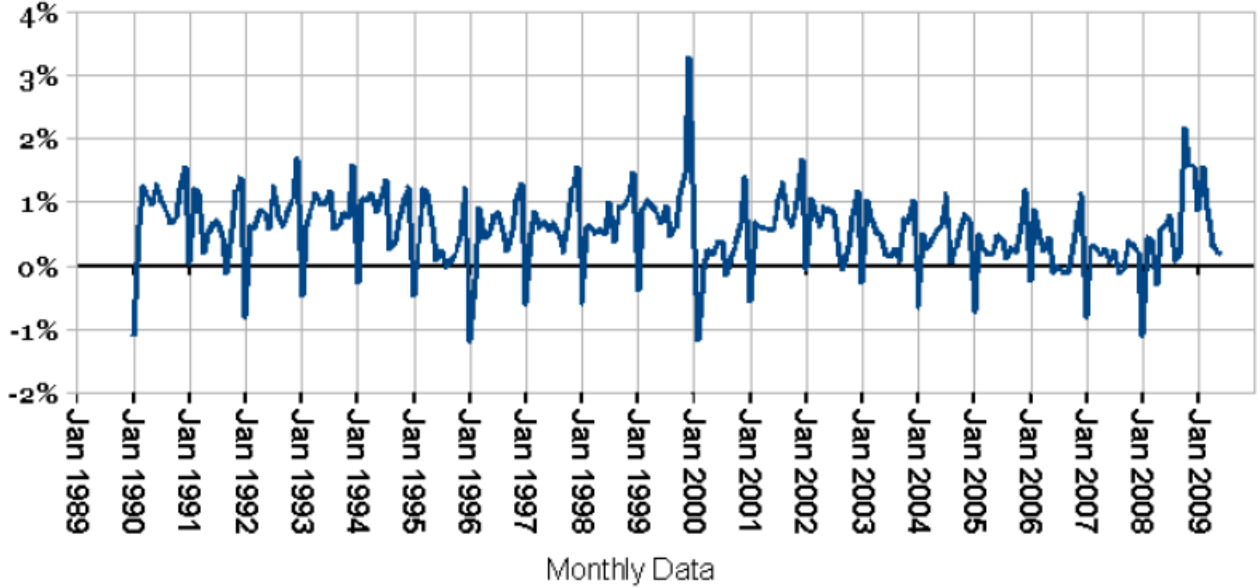

Figure 2. Monthly changes in currency component of U.S. money supply.

On the other hand, as it is known from the CAPM theory, the same risk-adjusted rate of return might be decomposed into the sum of riskless rate, $r$, and risk premium, $\lambda \sigma$:

(10) $\mu = r + \lambda \sigma$

where $\lambda$ is a unit risk price. From (9) and (10) it follows that difference between riskless rate and current yield is equal to the difference between capital gain and market value of risk:

(11) $r - \delta = a - \lambda \sigma,$

which will be used later. Altogether, equations (5-6) and (9-10) form basic structure of the stochastic money – debt model. These equations will be solved in the subsequent paragraphs in order to get simple functions of expected debt and new debt.

## Investors' motivation: expected and market debt

Debt, in the model, is assumed to have three different valuations simultaneously. The nominal debt value, $F$, is considered to be constant, being fully collaterized by real resources. The phenomenon of the complete debt collaterization by real resources takes place until some (critical) point of money issuance. On this interval the debt par measures the connectedness of financial and the real economic systems. On the other hand, the par debt, if being fully redeemed, is a measure of a riskless debt. The latter, as it will be shown later, is equal to the sum of its market value and the value of guarantees of financial stability by the society on behalf of the government and the central bank. Next, the market debt, $D(s_t)$, is assumed to be dependent upon money issuance, hence it is a risky one, with value being determined by the market forces of supply and demand. Market debt, quite naturally, cannot exceed its par value, $D(s_t) \leq F$. Contrary to that, the expected debt, $B(s_t)$, might take any value - larger, smaller or equal - to the par. It is treated by investors as perpetuity and depends upon liquidity issuance. This assumption, being justified by the fact that debt as a whole was never fully redeemed, simplifies all the model calculations.

The market debt value, $D(s_t)$, was supposed to have characteristics of an interest rate structurized product similar to a callable bond, or to its modern analogue – the exchange-traded note, ETN[6]. Remember that a non-callable bond could be decomposed into a callable bond and an embedded option to call the bond back were the issuer to consider such an operation to be profitable (it is, usually, when interest rates are going to decrease):

$$Noncallable\ bond = Callable\ bond + Call\ Premium$$

which reflects the fact that a callable bond has a smaller value to investors than a non-callable one (Levinson, 2010). Since investors have the right, but not obligation, to acquire new debt, the expected debt could be treated as an exercised callable bond which loses its callable feature:

$$B(s_t) = D(s_t) + f(s_t)$$

The investors' possibility to acquire new debt is represented (see Figure 3) in the model as an option to purchase new debt or a plain-vanilla call option:

(12)    $f(s_t) = [B(s_t) - D(s_t), 0]^+$,

being written on the expected debt value $B(s_t)$ with $D(s_t)$ as a strike price[7]. Quite naturally, intentions of investors to purchase new debt might be realized only, if they do possess of some amount of liquidity. Hence function $f(s_t)$ might have an important interpretation as an amount of credit available. The latter represents the credit amount as a result of the commercial banks activity that should be added to monetary base issued by the central bank in a manner analogous to calculation of money aggregates. The value of an option $f(s_t)$, being exercised, would represent the value of a new debt.

---

[6] The relatively new financial instrument – Exchange Traded Note – is a debt instrument invented by Barclay's Bank in 2006 (Wright et al, 2009).

[7] A callable bond ought to have a larger coupon and yield than the regular bond. For example, when the regular bond is $\frac{\$7}{0.07} = \$100$ it might be either $\frac{\$9}{0.09}$ or $\frac{\$8}{0.08}$. This example shows that the market value of a callable debt, depending upon its coupon differs from the ordinary one.

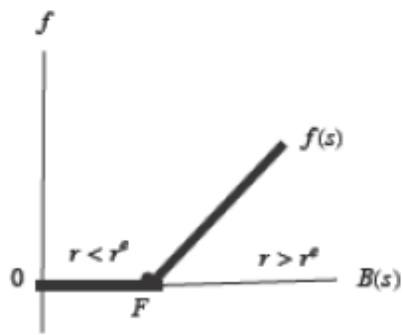

Figure 3. Matured option to purchase new debt.

The upfront option premium might be zero or very small. Usually buyers exercise their option to purchase new debt if its expected value exceeds the par, and do nothing otherwise. For example, anticipating increase in liquidity due to the central bank "easy money" policy they would reasonably try to benefit from interest rates decreases that make the expected debt larger than its par value. This pattern is formed by the typical behavior of investors who expect decreasing interest rates in the future or would perform the "flight to quality" under market distress and "quantitative easing". Since investors are not obligatory to make purchases, they could refuse to buy new debt in the case of interest rates increases. The above said assumption makes total debt similar to a callable bond. Hence their right to buy and increase their debt holdings should be embedded into the debt value, thus making rational investors to consider its expected value generally to be larger than the par.

For any market value of a debt, its expected value being stripped of the option to buy new debt becomes a simple structurized product of a following form:

(13) $B(s_t) = D(s_t) + [B(s_t) - D(s_t), 0]^+$.

Evidently, equation (13) in terms of liquidity issuance represents behavior of a rational investor that hinges around the expected interest rates as was depicted in (Keynes, 1936).

### Investors' motivation: the par and market debt

On the other hand, bond holders are concerned with the possibility of losses due to increasing interest rates, and, as financial analysis demonstrates, these losses are, usually, higher the smaller money issuance. By anticipating stochastic money flows in the future, bond holders could protect their wealth by acquiring various debt guarantees that are widely traded in financial markets. The major part of such guarantees in the modern financial systems are provided by the central bank which gives investors the free access to loans in the periods of financial distress, as well as by the state system of deposit insurance. It should be noted that while today's traditional banking system was made safe through the deposit insurance and liquidity provisions provided by the public sector, the shadow banking system – prior to the onset of the financial crisis of 2007-09 – was presumed to be safe due to liquidity and credit guarantees provided by the private sector.

These guarantees[8], being combined, form the so called put-to-default option (Poszar et al, 2010) that might be defined as

(14) $P(s_t) = [F - B(s_t), 0]^+$.

Looking from this angle, the debt market value, again, becomes a simple structurized product of the following form:

(15) $D(s_t) = F - [F - B(s_t), 0]^+$.

Since put and call options have the same strike price (and maturity profile) equations (14) and (16), being taken together, would describe investors' behavior via the "chooser" option which allows them to benefit from the large changes in the debt value. Being combined, these equations brought about the market value of aggregate debt, $D(s_t)$ (see Figure 4) that satisfies the following condition:

(16) $B(s_t) - f(s_t) = D(s_t) = F - P(s_t)$.

Equation (16) is nothing more than the model representation of the put-call equivalence theorem. It is fulfilled for any amount of money issuance in such a way that by increasing monetary base, the central bank supports the larger amounts of the credit available but, due to (17) that stochastic process is followed by the appropriate decrease of put-to-default option. Thus the financial system as a whole gradually, though randomly, evolves towards the state that could be described as a shadow banking system. The latter is associated with a very loosely regulated trading of financial instruments by independent investors. Their trading activity is not supported or insured since the put-to-default guarantees are substituted for larger amounts of liquidity available for different market participants. Inherently, such a system absorbs more risks than the conventional system of financial intermediation.

### Assets and liabilities

For the debt-money financial system, evidently, the sum of par and the new debt is equal to assets hold in the system. For maturing option contracts it is easy to show that, by adding and subtracting both put and call values from r.h.s. and l.h.s. of equalities (16), a following expression for the total financial assets, $A(s_t)$, would emerge:

(17) $B(s_t) + P(s_t) = A(s_t) = F + f(s_t)$.

Next, by subtracting (16) from (17) due to the basic accounting equation:

(18) $A(s_t) = D(s_t) + E(s_t)$,

it is possible to get a following definition of the total capital (equity) value:

(19) $E(s_t) = f(s_t) + P(s_t)$,

where $E(s_t)$ is the value of the owner's capital for financial system as a whole. Note, that due to the definition of the debt purchase option (12), the equity in the system (19) is different from its analogue in the well-known Merton model. Equation (19) implies that options, even not being exercised, do have positive value due to some additional abilities of investors provided by the well developed financial system. These possibilities are responsible for the Pigou effects upon the total wealth that might have disastrous consequences, if being exaggerated.

---

[8] On the microfinancial level this option is analogous to the Merton put-to-default option (Merton, 1992).

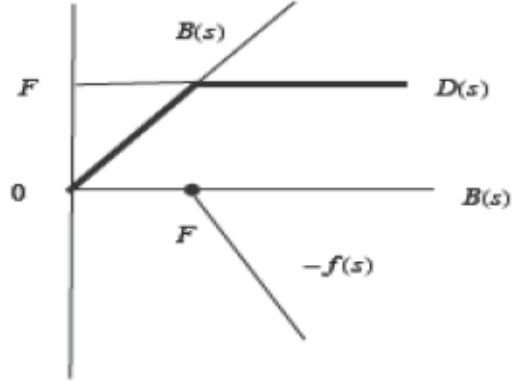

Figure 4. Market value of a debt at options maturity.

Next step in the model construction is to get a simple nonrandom representation of a debt as a function of money issuance.

## The expected debt valuation

In an uncertain financial market the aggregate debt is evolved stochastically due to random liquidity dynamics. Thus, transforming the debt infinitesimal change by applying the Ito lemma and substituting equation (6) into its expansion, we get the following stochastic equation:

(20) $dB = [as_t B'(s_t) + 0.5\sigma^2 s_t^2 B''(s_t)]dt + \sigma s_t B'(s_t)dz_t$

where debt derivatives exist for they are being taken with respect to liquidity issuance $s_t$. Coefficients of deterministic and random components in equations (5) and (20) could be equated, respectively, thus forming the following pairs:

(21) $\mu B(s_t) - s_t = as_t B'(s_t) + 0.5\sigma^2 s_t^2 B''(s_t)$

(22) $\sigma B(s_t) = \sigma s_t B'(s_t)$.

Via simple algebraic manipulations using (11), equations (21) and (22) could be transformed into the following inhomogeneous differential equation with respect to the debt function $B(s_t)$:

(23) $0.5\sigma^2 s_t^2 B(s_t)'' + (r - \delta)s_t B(s_t)' - rB(s_t) + s_t = 0$.

This second order equation is an ordinary analogue of the well known Black-Sholes partial differential equation, hence it has a much simpler, and intuitively understandable, solution (Dixit, Pindyck,1994). The solution to the homogeneous part of (23) could be found as a linear combination of power functions $B(s) = Bs^\beta$, while its inhomogeneous part has a solution as a linear function of money issuance, $\frac{1}{\delta}s_t$. Hence the expected debt function, $B(s_t)$, has the following representation:

(24) $B(s_t) = B_1 s_t^{\beta_1} + B_2 s_t^{\beta_2} + \frac{1}{\delta}s_t$,

where $\beta_1 < 0$ and $\beta_2 > 1$ are the real and distinct roots of a characteristic equation:

(25) $\quad 0.5\sigma^2 \beta(\beta - 1) + (r - \delta)\beta - r = 0$

that corresponds to the homogeneous part of (23). Since $\beta_1 < 0$ the first component of solution (24) for the very small money issuance goes to infinity. Hence, in order to preserve the economic sense of solution, the constant $B_1$ in (24) should be chosen as zero. This is so called "the absorption" condition requiring zero debt value in the absence of money issuance: its violation would lead to the unlimited debt growth that should be excluded from the model. The second constant in (24) is taken as zero, $B_2 = 0$, in accordance with the assumption of a debt as a never fully redeemed bond, or the perpetuity. Hence the expected debt value, $B(s_t)$, becomes the particular solution to equation (23) of the following form:

(27) $\quad B(s_t) = \frac{1}{\delta} s_t.$

As it follows from (27), the expected debt value is just the perpetual (capitalized) future stream of coupon payments being discounted by the current yield $\delta$. It should be noted that an anticipation of money growth makes rational investors to be complacent with the current yield, $\delta$, in spite of the fact, that it is smaller than the risk adjusted rate, $\mu$. This feature of the bond value (27) becomes evident after taking expectation of the random money issuance:

(28) $\quad B_0 \equiv \langle B_0 \rangle = \int_0^\infty \langle s_t \rangle \exp(-\mu\, t)\, dt = s_0 \int_0^\infty \exp[-(\mu - a)]\, dt = \frac{1}{\delta} s_0.$

According to (28), while investors do anticipate increases in the future money issuance, they should use precisely the risk adjusted rate, $\mu$, in order to discount flow of future payments properly. Hence evaluating (at point $t = 0$) the conditional expectation of the debt value brings about the same formula as in (27), assuming that, in the model, investors would consider the expected debt value as a simple perpetuity. For the financial system as a whole such an assumption does not contradict to the reality since the total debt (both public and private) should not be redeemed fully, contrary to individual debt of any single participant of the market. The total debt, being combined with anticipated future flow of liquidity, constitutes, as it appears, the persistent process of debt assets overvaluation that is known as "large asset price deviation". At the critical point of money issuance the asset price divergence reaches its maximum, as it will be shown later.

### Investors' new debt portfolio

In the model financial investors combine their decisions to buy the new and guarantee the existing debt. Thus their decisions are of hedging type, and it is assumed that the dominant group of investors is able to perform such an operation. In other words, under the normal regime any quantity of risk sellers would meet market participants taking the opposite position.

Let the incremental portfolio, $\Phi(s_t)$, consisting of money issuance and new debt, be represented as follows:

(29) $\quad \Phi(s_t) = \theta_1 s_t + \theta_2 f(s_t),$

where $\theta_1, \theta_2$ are the constant weights of new money and new debt, respectively. Since the new debt could be acquired due to credit, function $f(s_t)$ represents the amount of credit available to investors. This incremental portfolio could be made riskless, if investors are to choose special values of constants, namely, $\theta_1 = -f(s_t)'$ and $\theta_2 = 1$. With these weights, and due to equation of random money issuance (6), infinitesimal change $d\Phi$ to the incremental portfolio (29) becomes riskless:

$$(30) \quad d\widetilde{\Phi}(s_t) = 0.5\sigma^2 s_t^2 f''(s_t)dt.$$

For the shortest period of time $dt$ the return on portfolio (29) could be made riskless, if the hedged portfolio return is adjusted with the convenience yield of money, $\theta_1 \delta s_t dt$, lost due to hedging. This requirement gives rise to the following equation:

$$(31) \quad r[\theta_1 s_t + \theta_2 f(s_t)]dt = 0.5\sigma^2 s_t^2 f''(s_t)dt + \theta_1 \delta s_t dt.$$

Condition (31), modified by using the hedging values of constants and dividing through by $dt$, could be reduced to the following equation for the riskless portfolio held by investors:

$$(32) \quad 0.5\sigma^2 s_t^2 f''(s_t) + (r - \delta)s_t f'(s_t) - rf(s_t) = 0.$$

The important and rather unexpected result of these transformations is that the "new debt", or the credit, equation (32) has the same parameters (hence the same characteristic equation) as the homogeneous part of the debt value equation (23). Thus, by performing the same procedures of its solving, we get the value of an option to buy new debt as a function:

$$(33) \quad f(s_t) = K_1 s_t^{\beta_1} + K_2 s_t^{\beta_2}.$$

Due to the absorption condition the first constant in the r.h.s. of (33) has to be zero. Hence the option to purchase new debt (33) becomes the power function of a random liquidity $s_t$:

$$(34) \quad f(s_t) = K s_t^{\beta}$$

where $K \equiv K_1 > 0, \beta \equiv \beta_1 > 1$. Note that the value of the credit expansion is unrestricted. Hence mild credit issuance posits no problems (in the short run) but a persistent and unbounded credit expansion could create financial bubbles and crises.

Since, under the normal conditions, the call option should be exercised "in the money", investors, quite naturally, are motivated by the desire to maximize its value. That could be done if money issuance goes along the lines of "easy money" policy of the central bank. Such a policy, or the "quantitative easing, QE, in the modern parlance, facilitates important changes in the market structure.

### "Large asset prices deviation" at the critical point

The doctrine of the "large asset prices deviation" forms foundations of studying of the modern financial system (Rajan, 2005). The phenomenon of divergence market prices from fundamental value of assets could emerge in the model when investors, while continuing their strife for profit maximization, switch their orientation *en masse* from monitoring the market, $D(s_t)$, to the expected debt value, $B(s_t)$. In economic terms, such a behaviour of market participants might be described as follows.

The central bank, in a conduct of "easy money" (or the quantitative easing) policy, issues money according to (6) thus changing its quantity circulated in the system. Additional liquidity pumping by the bank into financial markets is anticipated by investors according to (8). Since, under the circumstances, interest rates are expected to decrease, investors are doing their best to benefit from the higher debt value. Hence, in accordance with (13), they buy new debt thus forming greater demand for it. The latter gives rise to a persistent asset price growth along (27). Adjusting to the changing environment, investors continuously rebalance their portfolios by eliminating money - major source of uncertainty - from their incremental portfolios (29). Investors exercise their options and maximize their value in order to buy new debt. Meanwhile, along with the easy money policy debt guarantees are going down due to increasing money issuance. Since maximization of the value of new debt purchases is supported by money issuance, investors' purchases of new debt are going on in the atmosphere of a widespread optimism. Asset prices are going up persistently and growing asset prices are "financial narcotics", as W. Baffett once said. All the features described above, appear to be typical characteristics of a growing financial bubble that is being formed by divergence of market prices from the value of financial assets.

Additional purchases of a new debt are induced by the liquidity increases up to its critical level at $s = s^*$, where asset price divergence reaches its maximum. It is impossible to estimate this point by standard methods, however. For example, the straightforward attempt to solve equation $D(s^*) = F$, upon substitution of functions (27) and (34) into this single equation, reveals two unknown parameters: $K, s^*$. Hence, in order to find the critical money issuance, the latter should be represented as a "trivial" solution to the dynamic programming problem (Dixit, Pindyck, 1998) in which the second order differential equation (32) has to be complemented with three boundary conditions. The boundary conditions include the initial value condition, $f(0) = 0$, together with the value-matching condition:

(32) $\quad f(s^*) = B(s^*) - F,$

and the smooth-pasting condition (in derivatives with respect to random variable $s_t$):

(33) $\quad f(s^*)' = B(s^*)'.$

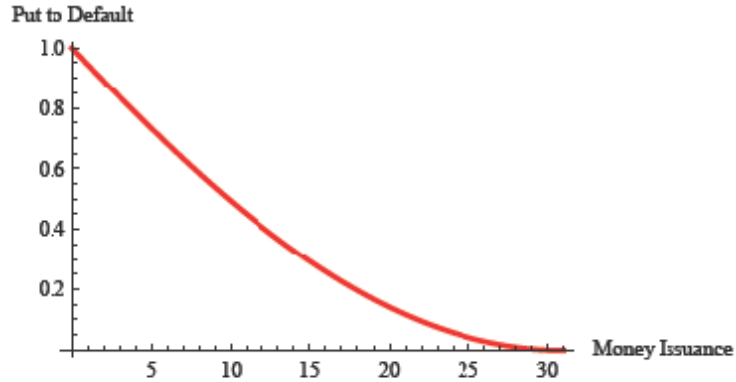

Figure 5. Value of the debt protection.

Upon substitution of the expected debt value (27) and the new debt value (34) into equations (35) and (36), point $s_t = s^*$ could be easily found as the following quantity:

(34) $\quad s^* = \dfrac{\beta}{\beta-1}\delta F.$

Money issuance at the free boundary point simultaneously maximizes the expected debt value, $B(s)$, which increases up to the amount of

(35) $\quad B(s^*) = \dfrac{\beta}{\beta-1}F.$

Thus, at the critical point $s = s^*$ of money issuance the central bank policy helps investors to deliver maximum value to the new debt, hence the debt value itself. It is evident from the put-call equivalence theorem (17) that it is possible if the put-to-default value would reach zero, $P(s^*) = 0$, while the market value of debt reaches its nominal value, $D(s^*) = F$.

The debt purchase option, $f(s^*)$ at the critical point appears to be "in the money" which suggests that it could be exercised. Whether it is exercised or not, depends, however, upon the broad market conditions that mark the important qualitative changes underwent in the system: at the critical point social guarantees of the market stability become substituted for the credit availability and their disappearance opens way to further market metamorphosis.

|  | **Margin account** | **iPath ETN** |
|---|---|---|
| $F$ | Own capital | Closing indicative note value |
| $f(s^*)$ | Borrowing "on margin" | The-then current financing level |
| $B(s^*)$ | The equity position | The long index amount |

The Table given above represents an interesting analogue between the debt structure at the critical point and the value of *iPath ETN* – structurized debt instrument invented by the Barclays Bank in 2006 (iPath, 2007). The analogy is based upon the technique of borrowing "on margin" which helps an investor to increase her/his long position beyond what can be established with the cash currently available within his/her account. For example, assuming we have such an instrument, then $B(s^*)$ could be associated with the long index amount, $F$ – with the CINV (closing indicative note value), and $f(s^*)$ - with the financing level. The analogy, discerned above, might be helpful, in our view, in the further studying of systemic characteristics of financial markets via analysis of their actual performance.

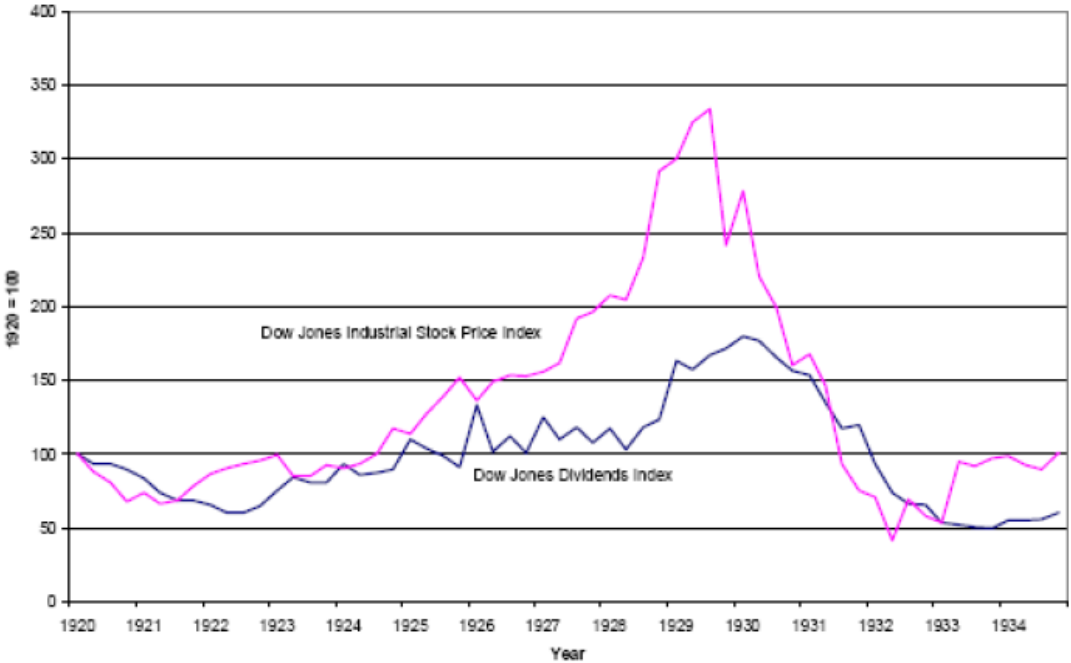

Figure 6. Bubble in the U.S. stock market.

**Critical point without herding**

To purchase new debt investors in aggregate have to spend their money hence converting the amount of credit available into new debt. Their demand supports the growing price of new debt thus inducing investors to substitute the market debt value for its expected value. The coherent actions of investors generate *en masse* the process of a persistent debt overvaluation. It follows from (16) that the expected debt exceeds its market value for any positive call option. Hence at the critical point $s = s^*$ the expected debt value increases at the rate of

(36) $\quad \frac{B(s^*)}{D(s^*)} = \frac{\beta}{\beta-1} > 1,$

which reflects the scale of the asset prices divergence at the critical point. Its magnitude corresponds to a short run Pigou effect of the total wealth increases under the normal regime of financial market. It should be noted that under some conditions this effect might be totally spurious, thus incurring losses of investors' wealth in the time of crisis.

Under "normal" conditions, as it follows from (17), the total assets value at the point $s^*$ is equal to:

(37) $\quad A(s^*) = F + f(s^*),$

since, by definition, the market debt value at the critical point equals to its nominal value, $D(s^*) = F$. What is the amount of financial equity in the system at the critical point? The answer depends whether at this point investors exercise the call option or not. In its turn, these actions are determined by the process of their herding, in other words, to what extend their behavior becomes irrational thus depriving them of the capability to make rational assessments and impose self-restrictions upon their behaviour.

If at the critical point of liquidity issuance $s^*$ no herding takes place, the market participants behave under the prevalence of rational motives. Rational investors would abstain from exercising their call option since their own capital would disappear in that case. Thus the motive of preservation of their own capital would mean that:

(38) $\quad E(s^*) = A(s^*) - D(s^*) = \frac{1}{\delta}\frac{\beta}{\beta-1}\delta F - F = \frac{1}{\beta-1}F = f(s^*) > 0.$

Equation (35) implies that investors while maximizing the amount of the new debt continue to evaluate their risks rationally, in other words, they are cautious enough as to keep part of their wealth as their own capital, $E(s^*)$. Consequently, under the "no herding" condition due to accumulation of equity by the market participants, the "distance-to-default" magnitude, or the "system survival" probability, at the critical point would amount to the quantity:

(39) $\quad \Pr[survival \equiv Distance - to - Default] = \frac{A(s*)-D(s*)}{A(s*)} = 1 - \frac{\beta-1}{\beta} = \frac{1}{\beta}.$

Alternatively, without herding of the debt purchasers, the probability of financial default, however large, would be strictly less than unity:

(40) $\quad 0 < \Pr[default] = \frac{D(s^*)}{A(s^*)} < 1.$

Since, by definition, $\Pr[default] = 1 - \Pr[survival]$, the default probability in the case of no herding is equal to the quantity:

(41) $\quad \Pr[default] = \frac{\beta-1}{\beta}.$

Such a scenario was investigated in (Smirnov, 2005) as a preferable, though not realized in practice, outcome of the government debt collapse in Russia in August 1998. It should be noted that coherent behavior of investors hedging simultaneously their portfolios, though being an optimal one, might bring about some unexpected consequences to the market performance. As (Chan et al, 2005) pointed out, the 1998 default on Russian government debt induced dramatic increases in market correlations. Instead of being negligibly small, as in normal times, they turned to plus one virtually overnight. That phenomenon the authors termed as the "phase lock-in". In other words, in the ongoing process of herding, the collective behavior of investors would bring about dramatic changes to the market leading ultimately to its collapse. This singular nature of herding might be captured, for example, via models of a percolation (Stauffer, 2001) or the chaotic behavior of financial investors (Ausloos et al, 2006) but these are different avenues of financial research that is not pursued in this paper.

**Herding at critical point**

The process of (irrational) herding in the model implies that every one of financial investors while maximizing the option to buy new debt, would substitute the market debt value, $D(s_t)$, for its expected value, $B(s_t)$. Increasing asset prices are followed by growing leverage, and this process was studied extensively, for example, in (Adrian and Shin, 2008). Though the scale of a bubble given by (32) is finite, it is easy to notice that the market leverage, $D(s)/E(s)$, is capable to grow indefinitely when $E(s) \to 0$ and $A(s^*) = B(s^*)$. If, at the critical point, $s_t = s^*$, the expected debt value becomes equal to the total assets value, then amount of capital in the system diminishes virtually to the zero:

(42)    $E(s^*) = A(s^*) - B(s^*) = 0.$

As a direct consequence of such a development, financial leverage ratio

(43)    $\frac{A(s^*)}{E(s^*)} \to \infty,$

would theoretically, as shown in Figure 5, increase indefinitely at the critical point.

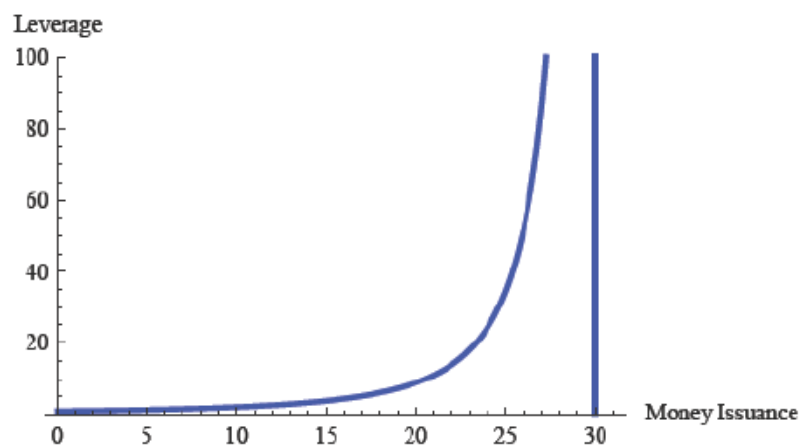

Figure 7. Leverage singularity.

Hence the ongoing process of herding among investors implies that *a posteriori* probability of default becomes equal to one:

(47) $\Pr[default] = \frac{B(s^*)}{A(s^*)} = 1$,

that makes the occurrence of a financial crisis to be a virtually inevitable event. From the economic point of view, it might be concluded that persistent increasing of the money issuance (excess liquidity) coupled with herding would lead to the systemic collapse. The latter is the sudden and dramatic decline in asset prices that might take place at the once benign point of money issuance $s = s^*$ were the herding features of investors behavior become pronounced enough to dominate the market. As it was stressed in (Rajan, 2005), "[the] prolonged deviations from fundamental value are possible because relatively few resources will be deployed to fight the herd". Unfortunately, few would want to go up against the trend that is originated by enormous mass of traders. Evidently, at the critical point investors trying desperately to get higher asset value are doomed to increase leverage as it had happened with investment banks on the eve of financial meltdown in 2007. Yet the collapse of the market at the critical point of money issuance, however probable, is not inevitable. The probability of a default at this point will be evaluated later, but now let us examine the system's behavior if the default is happened to be avoided.

Generally, origins of financial bubbles are hidden in the normal market since investors are allowed for the unrestricted exercising of their options that leads to a persistent substitution of the market debt value for its expected value. The crisis might have happened either at the critical point (under the hypothesis of herding) or later, at the point of the total collapse. Due to herding condition, the system might perform at the critical point dually. Without herding, investors would orient themselves around the market debt and prefer to accumulate own capital instead of new debt. These actions would decrease the probability of a default making it strictly less than unity. The herding process would have opposite consequences: to be consistent, investors are to use the expected, instead of the market, debt value as their guideline since the former serves as an adequate benchmark for the decision-making in a frenzy atmosphere of accelerating asset prices growth. Asset prices bifurcation at the Minsky point (to be discussed later) initiates the process of their divergence. At the critical point, where the price divergence reaches maximum, investors, all of a sudden, would realize that their own capital disappears while their assets are equal to the expected debt value. That awareness would have a devastating effect upon everybody's confidence in a system: market participants start to sell *en masse*, asset prices drop dramatically, and the system collapses. *A posteriori* probability of a total default (crisis) becomes equal to one while financial leverage starts to grow indefinitely, and the system collapses. Repeat again, that this phenomenon would have taken place at the critical point with some non-zero probability of default, and on a much greater scale it almost surely takes place on a next phase at point $\tilde{s}$.

<div align="center">**Total collapse of the system**</div>

If the central bank does not change its policy of money issuance (quantitative easing, for example) but investors impose self-restrictions refraining to exercise their call option, then financial system continues to evolve towards a shadow banking system. It is ultimately losing touch with its basic purpose - to transfer resources from ineffective to their effective usage. Instead, financial markets become, to a large extent, venues of a pure speculative activity. The unbounded credit expansion is accompanied with deprivation of the real wealth. In H. Minsky's parlance, the vast majority of investors become engaged in a Ponzi game facilitated by excess of money on a macrofinancial scale (Minsky, 2008). Remember that in the shadow banking system any counterparty is exposed to unprotected risks due to the absence of the put-to-default option. Ultimately, real resources serving as a collateral to financial assets, become negligible comparing with the huge amounts of new (uncollaterized) debt. This important qualitative change of financial system is captured via decreasing of the debt collateral, $F$, after the critical point of money issuance, $s^*$, due to the hectic acquisition of new debts by investors including banking institutions, being involved into the so called the "Ponzi game", using H. Minsky terminology. The Ponzi game of speculative activity goes on until the point $s = \tilde{s}$ where the market debt becomes completely worthless, $D(\tilde{s}) = F = 0$. At the point of money issuance, $s = \tilde{s}$, the complete absence of social guarantees of financial stability, or the zero put-to-default option, being combined with the worthless debt, gives rise to the system singularity. The general equation of the put-call equivalence (16) reduces to

(48) $\quad B(\tilde{s}_t) - f(\tilde{s}_t) = 0,$

meaning that the wealth of a nation becomes to consist of paper (or virtual) money only, uncollaterised by real resources. The wealth becomes completely worthless or the fool's gold. Under the circumstances, the collapse of the entire currency system is the only outcome of such an evolution. It was well understood and persuasively explained by the Post-Keynesian School (H. Minsky), and, in this important aspect, is completely in agreement with the major assertions of the Austrian School (Hayek, 2008). It should be noted, though, that in one important aspect the model differs from the Austrian constructions: it does not predict "a major currency crisis coming soon" as in interview of the "Austrian" economist P. Schiff (cited in Krugman, 2012) or similar doomsday predictions. though the timing of the latter is firmly tied up with the amount of money issuance.

As follows from the above said, different decisions of investors at the critical point of money issuance, $s = s^*$, are able only to postpone the inevitable financial crash. The credit crunch in the model either, with probability $p = 0.42$, takes place at point $s = s^*$ right away, or it comes later, at larger amount of money issuance, $s = \tilde{s}$, and "this time is different", indeed, since the probability of default is equal, invariably, to one. Thus, contrary to the soliloquy of Hamlet beginning with a famous assertion: "to be, or not to be, that is the question…" the investors dilemma, as formulated in the model – to buy or not to buy (an additional debt) – turned out to be, in fact, an illusory one[9]. In other words, the crisis would come inevitably, though, since at the critical point investors are able to buy some time, it might come either sooner or later. Yet, there is no escape from this vicious cyclical path without cardinal changes in the monetary policy. In this vital aspect, a cyclical behavior of the model dynamics literally repeats a famous L. vonMises assertion that was cited in the epigraph (vonMises, 1996, p.572).

**Natural cycle of the credit expansion**

The model permits to calculate different probabilities of default that could have happened under various amounts of money issuance. Their estimation follows the general approach to the default probabilities calculation (Kasapis, 2010; Hull, 2011). As was said in section 5, at the point $\hat{s}$ the value of a riskless debt is equal, by definition, to its market value plus the value of social guarantees:

$F = D(\hat{s}) + P(\hat{s}).$

---

[9] "To be, or not to be, that is the question:
Whether 'tis Nobler in the mind to suffer
The Slings and Arrows of outrageous Fortune,
Or to take Arms against a Sea of Troubles…"
*William Shakespeare, Hamlet, Act 3 scene 1*

Since put-to-default option could be represented as

(49) $P(\hat{s}) = p(\hat{s})F$

where $p(\hat{s})$ is the probability of system's default at that point. At this point probability to default is equal to

(50) $p(\hat{s}) = 1 - \frac{D(\hat{s})}{F}$ ,

while the par (riskless debt), $F$, could be expressed as

(51) $F = D(\hat{s})[1 - p(\hat{s})]^{-1}$.

Correspondingly, at the critical point $s^*$ since the debt value increases to $B(s^*)$ while at this amount it becomes risky, the following equation takes place

(52) $B(s^*) = F + p(s^*)B(s^*)$.

Thus the probability of default at the critical point is equal to

(53) $p(s^*) = 1 - \frac{F}{B(s^*)}$.

As it was said in the previous section, this probability a priory is strictly less than one, though it generally larger than default under the "normal" circumstances. It is easy to show that if

$$D(\hat{s}) \times B(s^*) = F^2 ,$$

then $p(\hat{s}) = p(s^*) = p$ and the probability of default could be calculated as

(54) $p = 1 - \omega^{-0.5}$ ,

where $\omega = \frac{B(s^*)}{D(\hat{s})}$ is the index of the debt growth. This probability was proposed in and evaluated for the debts of the European banking system (Smirnov, 2012).

Yet another important regularity in credit cycle might be discovered. The model shows that the longer credit expansion goes on, the higher becomes the probability of crisis:

(55) $p(\hat{s}) < p(s^*) < p(\tilde{s})$.

where, by definition, $p(\tilde{s}) = 1.0$. This regularity could be called *the natural credit expansion* enequality. It follows directly from inequality

$$D(\hat{s}) \times B(s^*) < F^2$$

since we have to compare probabilities to default at points of monetary issuance $\hat{s}$ $and$ $s^*$.

Interestingly enough, the footprints of the same idea could be found in a recent report of Bloomberg (Bloomberg, September 6, 2012). The news agency cited an essay of the London-based analytical firm SLJ Macro Partners LLP in which the authors, supposedly, came to the idea of "natural cycles" empirically, after a thorough analysis of the family of yield curves.

## General structure of a credit cycle

Possibility of excluding risks via hedging that helps to transform the system into a simple deterministic model, would open way to the analysis of a neatly formed structure of a financial cycle. First, deterministic behavior of a system along the money issuance axis reveals the existence of three, clearly distinguishable, points which define and separate qualitatively different regimes that would emerge in the process of money issuance. Second, critical points and different regimes of a market, in their turn, could be quite naturally identified as the Minsky phases of normal investing, speculation and Ponzi game (Minsky, 2008). Being studied along these lines, the model would demonstrate a striking similarity between sequential Minsky phases and a cyclical behavior of a financial system in the continuous process of credit expansion, as described by the Austrian School of Thought.

The phased behavior of a financial system is represented in Table 1. Repeat again, that its simplicity is noticeable in terms of money issuance only, while as a variable of time it, most likely, would possess a multifractal structure. For example, the fact that in 2008 the Fed balance sheet was increased more than two-fold (from \$0.9 billion to \$2.3 billion) for a period of time less than half a year, could have had an interpretation of a multifractal structure of a time variable: $[0, \hat{s}] \to t_1 \gg t_2 \leftarrow [s^*, \tilde{s}]$.

**Table 1.  General structure of a credit cycle**

| Money Issuance, $s$ | Debt Face Value, $F$ | Debt Expected Value, $B(s)$ | Debt Market Value, $D(s)$ | New Debt Value, $f(s)$ | Social Guarantees, $P(s)$ |
|---|---|---|---|---|---|
| $[0, \hat{s})$ | $F$ | up | up | up | down |
| $\hat{s}$ | $F$ | $B(\hat{s}) = F$ | $D(\hat{s}) > 0$ | $f(\hat{s}) = P(\hat{s})$ | $P(\hat{s}) = f(\hat{s})$ |
| $[\hat{s}, s^*)$ | $F$ | up | up | up | down |
| $s^*$ | $F = $ $= B(s^*)$ $- f(s^*)$ | $B(s^*)$ $= F + f(s^*)$ | $D(s^*) = F$ | $f(s^*)$ $= B(s^*) - F$ | $P(s^*) = 0$ |
| $[s^*, \tilde{s})$ | $F = D(s)$ | up | down | up | $P(s^*) = 0$ |
| $\tilde{s}$ | $F = 0$ | $B(\tilde{s}) = f(\tilde{s})$ | $D(\tilde{s}) = 0$ | $f(\tilde{s}) = B(\tilde{s})$ | $P(s^*) = 0$ |

The Minsky phases are characterized, generally, by different roots of equation (16) which, for convenience, is reproduced here:

$B(s_t) - f(s_t) = D(s_t) = F - P(s_t).$

In terms of money issuance, $s$, there exist three important critical points:

*no-arbitrage or equilibrium point*, $\hat{s}$, where investors are indifferent whether to possess expected or par value of a debt, $B(\hat{s}) = F$. Their indifference is due to their ability to swap put-to-default option for new debt, or vice versa, $f(\hat{s}) - P(\hat{s}) = 0$;

*the second critical point*, $s^*$, where investors maximize the value of the acquired new debt, $f(s^*) = \max_s f(s)$. Hence the equality appears: $B(s^*) = F + f(s^*)$ which takes place, if $D(s^*) = F$, $P(s^*) = 0$.

*total default (the crisis) point*, $\tilde{s}$, where the face value of a debt is completely lost due to virtual disappearance of its real collateral, $D(\tilde{s}) = F = 0$. The latter takes place due to excessive un-collaterized borrowing by financial institutions. Under these circumstances investors' wealth, consisting only of credit, $B(\tilde{s}) = f(\tilde{s})$ became, in fact, a fool's gold, and the entire financial system is doomed to collapse.

The normal regime of financial investing could be defined on the interval of money issuance $s \in [0, \hat{s}]$. Along this interval the central bank facilitates expansionary policy that is followed by increases in expected and the market value of a debt. Strictly speaking, financial calamities could have taken place within this semi-interval of money issuance, if social guarantees are absent or insufficient. Historically, something like that had taken place in 1998 in Russia due to restrictions on its money issuance being imposed by the IMF (Smirnov, 1999). Such a possibility is reflected in the model via the probability of default which at this point, as it was shown, is strictly less than zero.

The first signal of irreversible changes that are to happen in the system, might be discerned at the point $s = \hat{s}$ where the expected debt value becomes equal to its collateral: $B(\hat{s}) = F$. At this point investors are prepared to make a swap between the option to buy new debt and the put-to-default option since

$f(\hat{s}) - P(\hat{s}) = 0$.

After this point, the expansionary money policy of the central bank would motivate investors to exercise their option to buy new debt because its expected value becomes persistently larger than that of the put-to-default option. The expected debt being exercised loses its callable feature and persistently overvalues its market value. Hence the debt market value starts to diverge noticeably from its fundamental value.

At the critical point, $s_t = s^*$, financial system underwent its first major structural change: the market debt value reaches its maximum, $D(s^*) = F$, bounded by its riskless amount, while the expected debt became maximal, $B(s^*) = F + f(s^*)$. This is possible, however, due to the zero value of the put-to-default option, $P(s^*) = 0$. Thus the normal regime comes to an end, reflecting the final stage of a gradual process of substitution of standard banking intermediation for the direct trading in financial instruments. Standard banking intermediation has finally transformed into the shadow banking system. At the critical point $s^*$, the system may (or may not) come to a crisis with some non-zero probability. If the crisis is to take place at this point, it would be of a relatively small amplitude associated usually with the market correction.

If the central bank does not change its policy of money issuance financial system continues to evolve as a shadow banking system. The unbounded credit expansion facilitates a Ponzi on a macrofinancial scale. The random liquidity issuance being coupled with herding produces a kind of "irrational exuberance" among investors, and their collective behavior drives the system towards its total collapse. Ultimately, real resources serving as a collateral to financial assets, would become negligible in comparison with the enormous amount of a new, and uncollaterized, debt. Thus, at some point of money issuance, $s = \tilde{s}$, a complete disappearance of social guarantees of a financial stability, being combined with the worthless debt, gives rise to the system singularity. Under the circumstances, the collapse of the entire currency system might be conceived as the only outcome of a monetary evolution that, by definition, would take place with the unit probability of default.

## The excess of money accumulation

The sum of money issuance and funds realized as a new debt value, $s_t + f(s_t)$, reflexes a trade-off between liquidity and returns that investors usually face on financial markets. Cash in the pocket is perfectly liquid but earns no return while investment into new debt promises some pay-off, if being successful, but not liquid.

The system behavior at the critical point $s = s^*$ is a very misleading one. Though it was shown previously that investors maximize their acquisition of a new debt, their actions are far from being the most important factor of the system transformation. Actually, at this point the system, as a whole, accumulates excessive cash, $s$, as the means of exchange opposite to loanable funds, $B(s) - s$, and that means the formation of excessive money. Since mathematical structure of the model could have produced the entirely pervert impression about the system behavior, it is in need of some clarifications. For this purpose it would be helpful to compare investors' acquisition of debts with the straightforward re-lending process. According to the standard re-lending model, at the point $s_t = \hat{s}$, the banking system's assets, $\hat{s} + (1 - \delta)F$, and liabilities, $F$, would form a very simple configuration given by the following balance sheet:

| Assets | Liabilities |
|---|---|
| money (issuance), $\hat{s}$ | Value of riskless debt, $F$ |
| credit amount, $(1 - \delta)F$ | |
| Total: $F$ | Total: $F$ |

which could be represented as a standard accounting equation:

(56) $\quad \hat{s} + (1 - \delta)F = F \quad$ since $\quad \hat{s} = \delta F$.

According to (56), financial (banking) system as a whole has assets consisting of money, $\hat{s}$, and credit, $(1 - \delta)F$, while its liabilities are represented as riskless value, $F$, of resources deposited by the real system. In the linear re-lending process the money multiplier is given in by parameter $1/\delta$, namely, as amount of credit per unit of money issuance.

The same configuration is repeated when the system arrives at point $s^*$:

(57) $\quad s^* + (1 - \delta)\frac{\beta}{\beta-1}F = \frac{\beta}{\beta-1}F,$

with the only difference that when the system moved from point $\hat{s}$ to point $s^*$, its liabilities, hence money issuance increased in proportion of $\frac{\beta}{\beta-1}$.

| Assets | Liabilities |
|---|---|
| money (issuance), $s^* = \frac{\beta}{\beta-1}\delta F$ | Value of expected debt, $\frac{\beta}{\beta-1}F$ |
| credit amount, $(1 - \delta)\frac{\beta}{\beta-1} F$ | |
| Total: $B(s^*)$ | Total: $B(s^*)$ |

On the other hand, the new debt maximization forms the final configuration of a system at point $s^*$, as following:

(58) $\quad F + f(s^*) = B(s^*) \equiv \dfrac{\beta}{\beta-1} F.$

| Assets | Liabilities |
|---|---|
| money (issuance), $s^* = \dfrac{\beta}{\beta-1} \delta F$ | Value of riskless debt, $F$ |
| credit amount, $(1-\delta)\dfrac{\beta}{\beta-1} F$ | Value of the new debt, $f(s^*) = \dfrac{1}{\beta-1} F$ |
| Total: $\quad B(s^*)$ | Total: $\quad B(s^*)$ |

Thus, as it follows from (57) and (58), the system as a whole accumulates excess money in the amount of

(59) $\quad (1-\delta)\dfrac{\beta}{\beta-1} F - f(s^*) = [(1-\delta)\beta - 1]\dfrac{F}{\beta-1}.$

The huge amount of excess money facilitates the subsequent credit expansion leading to the ultimate collapse of the entire currency system. From the previous it follows that inequality

(60) $\quad \beta(1-\delta) > 1$

quantifies the vonMises assertion taken as an epigraph to the paper. Since in the model, as it will be shown later, $\beta \gg 1$, it explains the phenomenon of the crisis inevitability and deserves to be named as *the vonMises inequality*.

Finally, in the phase of the Ponzi game, the financial system would accumulate debt in such enormous amount that would dwarf its real collateral. Hence the system configuration would become a totally biased one supposing its collapse at point $\tilde{s}$.

| Assets | Liabilities |
|---|---|
| money (issuance), $\tilde{s}$ | |
| credit amount, $\left(\dfrac{1}{\delta} - 1\right) \tilde{s}$ | Value of the new debt, $f(\tilde{s})$ |
| Total: $\quad B(\tilde{s})$ | Total: $\quad B(\tilde{s})$ |

For the purpose of comparison on Figure 6 are given empirical data of excess liquidity in the eurozone banking system in 2009-11.

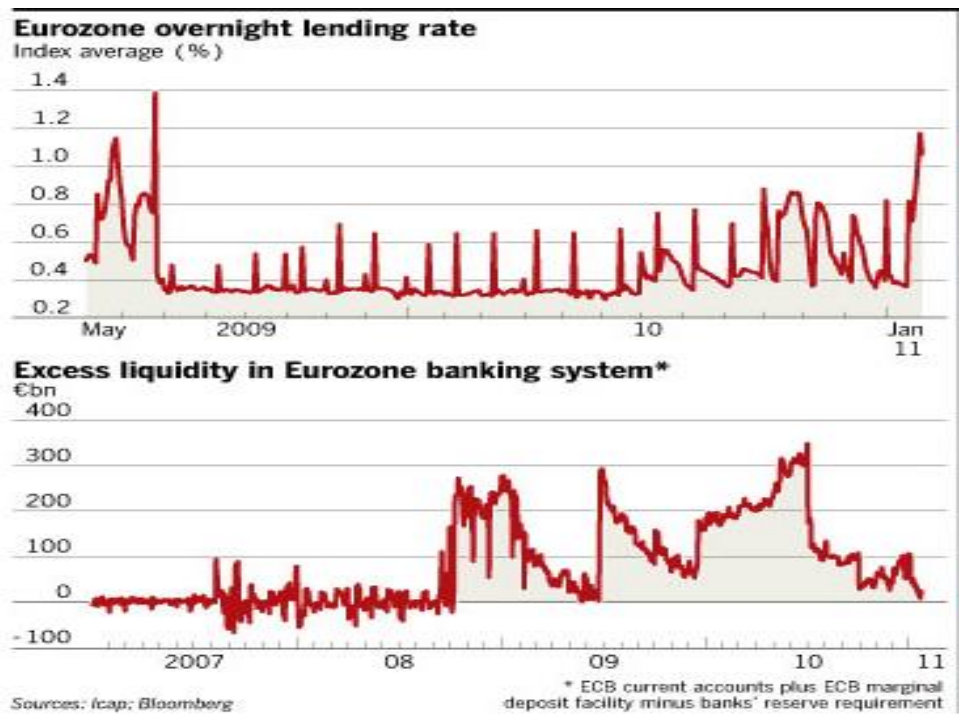

Figure 8. (From FT, January 31, 2011)

Remember, that in the model money does not play in full its function as "means of payment". It makes finance as a kind of a circular system, where debts are monetized within the system itself. To pursue such an approach, in spite of its artificial character, would seem to be helpful in understanding some prominent features of the modern finance. Namely, though the G4 economies central banks have dramatically increased their balance sheets, the impact of these actions upon the aggregate money demand was almost negligible. Thus, the excess money is doomed to lose its purchasing power which, if happens, would pave way towards the imminent collapse of the whole system.

**The Minsky point**

The concept of asset prices divergence implies the importance of discerning a bifurcation point between the trajectories of "normal" debt value given by function $D(s_t)$, and the overvalued debt trajectory of $B(s_t)$. This problem has an important real life equivalent associated with the early detection of a financial bubble. Economists debated this issue for a long time: it is enough to recall the critique of Greenspan's policy by P. Krugman for its inability to prevent the bubble growth in the housing and credit markets (Krugman, 2008). The point of bifurcation deserves to be called " the Minsky" point since it is the origin of all the subsequent troubles in the financial market[10]. As learned from the history of finance, the failure to detect properly the Minsky point would have had ominous consequences for the market for the avalanche of debts becomes virtually irreversible after it has been passed.

In the model, the asset price divergence becomes easily recognizable after the system passage through the point of bifurcation where the financial bubble emerges. It is reasonable to identify the Minsky point with intersection between either of trajectories $D(s_t)$ or $B(s_t)$ with the trajectory of the put-to-default, $P(s_t)$. Before the point

---

[10] In economic literature the "Minsky moment" is associated with the market meltdown (Cassidy, 2010).

of bifurcation $s_m$ the market or expected debt is fully guaranteed while it is not after. Thus, assuming the equality

$$D(s_m) = P(s_m)$$

takes place, the Minsky point $s_t = s_m$ could be found as the solution to

(61) $F = 2[B(s_m) - f(s_m)].$

On the other hand, for the condition

$$B(s_m) = P(s_m)$$

The Minsky point could be found as a solution to

(62) $F + f(s_m) = 2 B(s_m).$

Evidently, since the option to buy new debt is out of the money for the small liquidity issuance, the discrepancy between (61) and (62) would be rather small.

## Numerical primer

The model described above could be illustrated numerically as follows. The system parametrization was based on a rather realistic figure of a nominal debt in 200 trillion of dollars, $F = \$200tr$, that roughly corresponds to the amount of world debt as given by the IMF (GFSR, 2011). The world liquidity was estimated in amount of $9.6 trillion that approximated the amount of the world monetary base. Rates of return in the model simulation were taken as follows: the riskless rate was about 5 percent, $r = 0.05$, and annual risk-adjusted interest rate was equal to 7 percent, $\mu = 0.07$, per annum. The latter, equivalently, is equal to the sum of current yield, $\delta = 0.045$, and expected annual capital gain, $a = 0.025$. The quantity of risks (per annum) in the system was measured by annual volatility, $\sigma = 0.15$, which had its price per unit, $\lambda = 0.45$.

The characteristic equation of such a system:

$$0.5 \times 0.15^2 \beta(\beta - 1) + (0.05 - 0.045)\beta - 0.05 = 0$$

has two distinct real roots: $\beta_1 = -0.099$, and $\beta_2 = 2.404 > 1$, of which only the positive root has an economic meaning. The graph of that characteristic equation is shown in Figure 9. It is interesting to note that financial process in the model is of a fractal nature with the exponent being not too far from the so called "cubic law" (Lux, 2006).

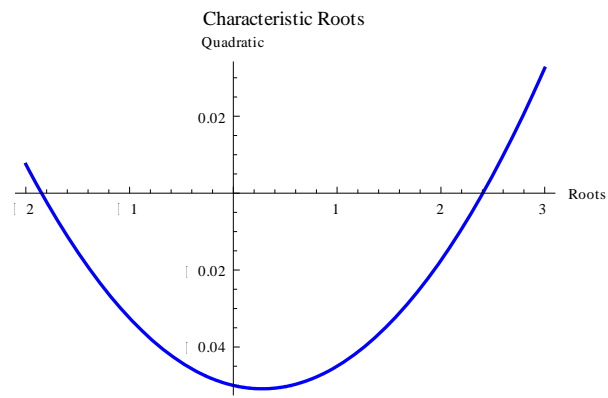

Characteristic Roots
Quadratic

Figure 9. Characteristic equation and its roots.

The major characteristics of the system cyclical behavior are summarized in Table 2. Its entries were calculated using the appropriate formulas of the model.

**Table 2. (in trillion of dollars)**

| Money Issuance, $s$ | Market debt value, $D(s)$ | Expected Debt Value, $B(s)$ | New Debt Value, $f(s)$ | Social Guarantees, $P(s)$ | Probability of Default |
|---|---|---|---|---|---|
| $\hat{s} = 9.0$ | 161.0 | 200.0 | 39.0 | 39.0 | 0.195 |
| $s = 12.4$ | | | | | 0.315 |
| $s^* = 15.5$ | 200.0 | 342.9 | 142.9 | 0 | 0.417 |
| $\tilde{s} = 28.9$ | 0 | 642.6 | 642.6 | 0 | 1.0 |

For example, in our numerical primer the critical point of money issuance $s^*$ is equal to

$$s^* = \frac{2.4}{1.4} \times 0.045 \times 200 = \$15.5 \ tr.$$

This quantity defines the expected value of a debt at the critical point:

$$B(s^*) = \frac{1}{0.045} \times 15.5 = \$342.9 \ tr,$$

and, correspondingly, the value of new debt:

$$f(s^*) = 0.2 \times 15.5^{2.4} = \$142.3 \ tr$$

where constant $K = 0.2$. Hence at the critical point $s^* = \$15.5 \ tr$, the expected value of debt equals to $\$ 342.9 \ tr$. The latter amount is the sum of the nominal debt $\$ 200 \ tr$ plus new debt $\$142.9 \ tr$.

The Minsky point, $s_t = s_m$ is defined by as the equality of debt guarantees to either the market or to the expected debt value. Hence this point is estimated as either $\$4.9 \ tr$ or $\$4.7 \ tr$, respectively. Note, that starting at this point, rational behavior of investors becomes less rational since they follow the trajectory of the debt overvaluation. The rational motivation prevails until the point $\hat{s}$ after which it turns out into speculation. This process becomes autocatalytic (the Ponzi game in the Minsky terminology) after the critical point $s^*$, and eventually ends up at point $\tilde{s}$ in total crisis. The model reveals an important feature of a crisis situation at the critical point. Even if private investors due to self-imposed restrictions upon the new debt acquisition avoid the crisis, the central bank continues to increase liquidity further. Such a behavior, incidentally, is fully in line with the monetary policy of quantitative easing after the current financial meltdown. Autocatalitic character is reflected via the disappearance of the real collateral to the debt. Phases of speculation and the Ponzi game together form the emergence of a financial bubble that could burst either at the critical point $s = s^*$ or, ultimately at $\hat{s}$. The overall picture of a financial cycle is given in Figure 8.

**Table 3. Probabilities to default.**

| Money issuance, $s$ (\$ tr) | Default probability, $p(s)$ |
|---|---|
| 9.0 | 0.195 |
| 12.4 | 0.315 |
| 15.5 | 0.417 |

| 28.9 | 1.0 |

The numerical primer supports the idea of the "natural cycle" in terms of probabilities to default which are given in Table 3. Evidently, larger money issuance in the system leads to larger probability of its default. The invariant credit expansion in the model is doomed to end up in the total collapse of the entire monetary system that would take place with probability 1.0.

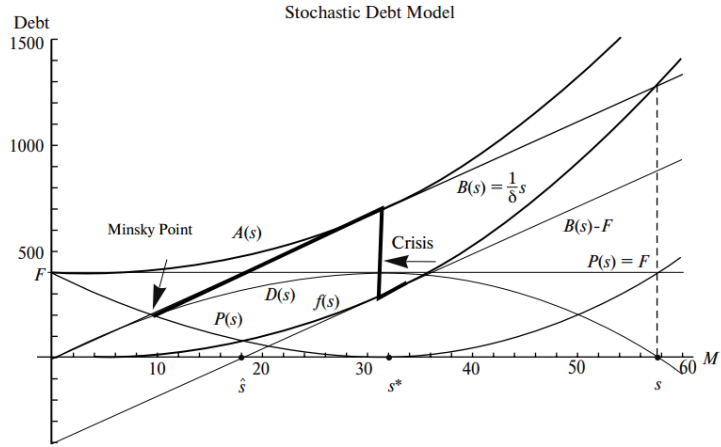

Figure 10. The Minsky phases in finance.

After the point of total collapse $\tilde{s}$, the cycle would repeat itself. The next, "rational" phase of a cycle could be conceived as deleveraging along the lines of the debt-deflation spiral elaborated by I. Fisher (Fisher, 1933). The behavior of investors is very cautious after that painful period: they prefer to form and accumulate debt guarantees in the first place thus making safer the post-crisis system. Yet, as investors' confidence gradually grows along the safer investment activity so would intensify their arrogance. Phases of speculation and the Ponzi game are clearly discernible along the same lines as thoroughly studied by H. Minsky. Without the doomsday predictions the model outcome of a credit expansion conforms, in general, assertions of the Austrian School, too.

**A temporal view of a system**

The system singularity which is followed out of zero equity condition at the critical point, $E(s^*) = 0$, might be deduced alternatively, via investors' expectations. Remind, that according to equation (1) at any point of time the sum of expected values for money and debt is given by

(63) $\langle A(t) \rangle = \langle M(t) \rangle + \langle B(t) \rangle$ .

The expected money aggregate, $\langle M(t) \rangle$, at time $t$ is equal, by definition, to

(64) $\langle M(t) \rangle = \int_0^t \langle s_u \rangle du = \int_0^t s_0 \exp[a\, u]\, du = \frac{s_0}{a}(\exp[at] - 1)$

due to expected money issuance given by equation (8). At the same time, by taking the expected debt value from (27) we get

(65) $\langle B(t) \rangle = \frac{1}{\delta} \langle s_t \rangle = \frac{1}{\delta} s_0 \exp[a\, t]$.

Hence, by adding (64) and (65), the expected asset value at time $t$ becomes to be expressed as follows:

(66) $\langle A(t) \rangle = \frac{s_0}{a}(\frac{\mu}{\delta}\exp[at] - 1)$.

The time of a crisis as "time zero", $t^* = 0$, would correspond to the critical point of money issuance, $s^* = s_0$, and the total asset value (66) would equal to

(67) $\langle A(t^*) \rangle = \langle A(s^*) \rangle = \langle B(s^*) \rangle$.

Hence, analysis of investors' expectations leads to the same result as before: the value of total assets at the critical point (due to herding) consists of expected debt only. As such, the result (67) would have suggested the ergodic character of financial processes, though this assertion is in need of the further exploration.

Yet one more important comment should be made with regard to the "absolute" quantities of debt, money and value of total assets. Comparing equation (1) being evaluated at the time of a crisis $t^* = 0$

(68) $\langle A(t^*) \rangle = \langle M(t^*) \rangle + \langle B(t^*) \rangle$

with equation (17) evaluated at the critical point $s = s^*$

(69) $A(s^*) = B(s^*) + P(s^*)$

we have to conclude that

(70) $\langle M(t^*) \rangle = P(s^*) = 0$

since $P(s^*) = 0$ due to (35). Since total money aggregate is a nonzero quantity, equality (70) being taken literally, would have formed a logical contradiction, though it is not. In fact, this controversy is a spurious one, and could be explained as a manifestation of the system singularity. It is a well known fact that any financial crisis is a catastrophic shortage of liquidity. Hence equality $\langle M(t^*) \rangle = P(s^*) = 0$ just demonstrates the absence of money at the critical point or, which is the same, its negligibly small quantity comparing to the total debt value. Thus, while the random quantity of money is not zero at the critical point, the expected money aggregate is zero. Financial system devastated by the crisis becomes singular at the critical point which is manifested by the "zero-money" condition.

## Percolation and bubble singularity

Historically, financial bubbles were always precursors of crises (Friedman, 1963; Kindleberger, 2000). Bursting bubble, in its turn, could be represented formally via debt singularity that appears due to herding. Singularity takes place for the systems of the infinite dimension while empirically all the systems are of finite dimensionality. Such a contradiction, well known in the natural sciences, manifests itself in the instable characteristic scale while samples are increasing.

The model suggests the existence of the finite, however large, amount of the debt, $B(s^*)$, at the critical point of money issuance, $s_t = s^*$. This feature, empirically quite correct since in reality debt outstanding is always a finite amount, theoretically could be generalized to embrace the theoretical concept of singularity as a solution to a non-linear equation. Looking at this angle, linear function $B(s_t)$ serves as a poor representation of asset prices dynamics near the critical point since the latter, being chosen by the herd of investors, is a highly nonlinear one. Hence, formally, it seems to be more preferable to represent asset prices dynamics near the critical point as a singular process. It implies that the essence of herding is not only in the asset prices overvaluation but in the transformation of investors' behavior into a highly nonlinear process of autocatalytic type result ing in the new debt acquisition. The pronounced mimicry of investors forces the debt value to increase, accelerating this process simultaneously. The asset prices start to behave both in nonlinear and almost deterministic manner. The stock quotations become a commonplace: the story goes that in 1929 Joe Kennedy (the father of the future US President) liquidated his portfolio when he heard that a shoeshine boy was giving stock tips repeatedly. The trajectory of the blown asset prices becomes the only one along which the singularity might take place. Since the "normal" debt reimbursement is implied by finite amount of the par debt, in the process of herding going on in the small neighborhood of the critical point investors would completely ignore such a possibility. The process of acquiring new debt becomes irrational once investors dampens the mere notion of a fair price of an asset.

Irrational bubbles occur when investors develop an enthusiasm for particular class of assets like stocks in the late 1990ties or houses in the beginning of 2000ties[11]. Quickly blowing financial bubbles could be studied via models of financial percolation. Percolation is a huge body of knowledge with a large spectrum of applications from physics to chemistry, to earthquakes to avalanches to forest fires (Stauffer, 2009). In finance percolation models are useful in describing interactions of investors via geometric configurations of sites being formed randomly on a large 2D grid. Monte Carlo simulation of percolation models shows that in the vicinity of a critical point these interactions might lead to formation of a huge spanning cluster of sites that transforms the quality of the financial system. The latter is due to a sudden increase of the "connectedness" among the hitherto independent financial investors (Smirnov, 2010). It follows that near the critical point financial bubble starts to expand in a highly nonlinear manner, probably first noticed by J.M. Keynes in his description of "speculation" and "enterprise" in financial market (Keynes, 1936). The model demonstrates that since investors acquire new debt unboundedly, the total debt value at the critical point becomes infinite, and the bubble bursts very quickly.

Taking these considerations into account, the new debt function (34) is to be modified to follow the singular process. The simplest modification of this sort could be proposed as the following:

(71) $f(s_t) = K s_t^{\beta} + h * (s^* - s_t)^{-\gamma}$ ,

where the herding parameter is

$h = \begin{cases} 1, \text{if herding}; \\ 0, \text{if no herding}; \end{cases}$

and $\gamma = 2.39$ is one of the percolation invariant constants (Stauffer, 2009). As it might be seen from Figure 11, herding modifies the new debt function (34) significantly only in the small neighborhood of the critical point $s^*$.

---

[11] There is a long tradition to study rational financial bubbles as well (Lux, Sornette, 2002)

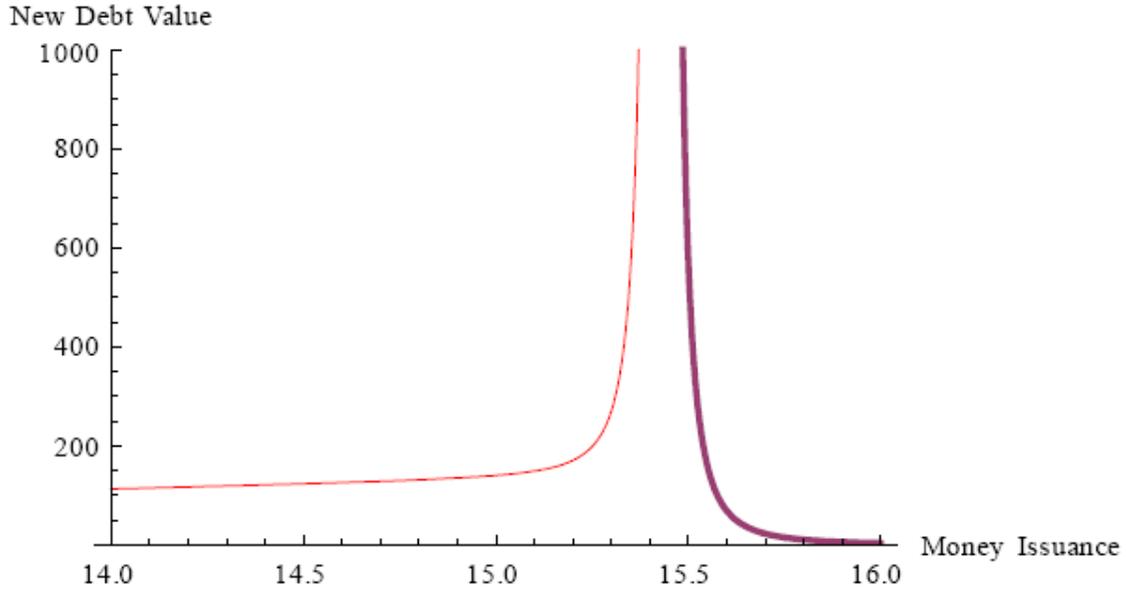

Figure 11. The new debt function singularity.

In our numerical example the new debt function was taken as

$$f(s_t) = 0.2\, s_t^{2.4} + h * (15.43 - s_t)^{-2.39}.$$

which, in the vicinity of the critical point, is dominated entirely by its second component. Ignoring the first component in (61) and differentiating it through by money issuance, we get

(72) $\quad \frac{df}{ds} \sim (s^* - s_t)^{-\gamma - 1}$

where (~) is the sign of asymptotic equality. According to (72), in the small neighborhood of the critical point, the process of herding accelerates tremendously the debt accumulation that quickly leads to its singularity.

It follows that near the critical point $s = s^*$ financial bubble starts to expand in a highly nonlinear manner which bursts very quickly as represented in Figure 9. Repeat again that the same process, but on a much larger scale, is going on at the point $s = \tilde{s}$. Being stimulated by herding investors acquire all new debt unboundedly, hence the total debt value at the critical point becomes infinite. In reality that signifies the burst of a bubble defined as a system singularity at the critical point. Looking at the different angle, however large amount of money becomes, in fact, negligibly small comparing to the infinitely large debt value. The shortage of liquidity which is a financial crisis *per ce,* is a result of the eventual bursting of a financial bubble that takes place at the critical point of money issuance. To overcome the consequences of a crisis, the money issuance in the model should be increased even further than before the crisis. That is precisely what had been done by major central banks in the aftermath of credit crunch 2007-09.

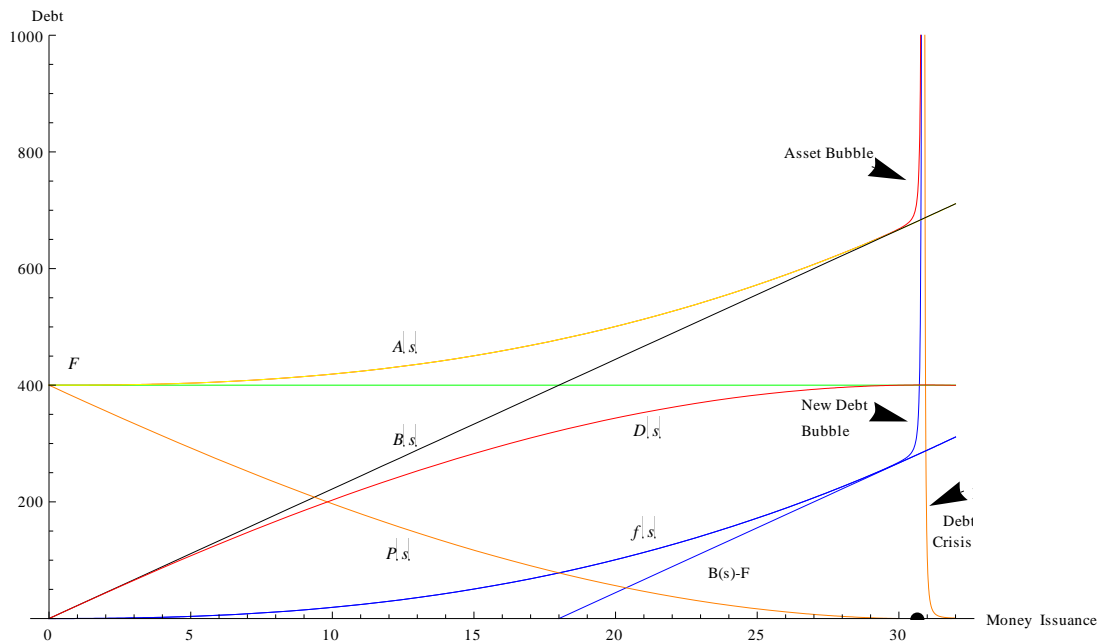

Figure 12. Financial bubble and crisis.

The trajectory of the blown bubble becomes the only one, and asset prices increase along it in nonlinear and almost deterministic manner. In the process of herding investors, quite in accordance with "the greater fool theory", completely ignore possibilities of "normal" reimbursement of the par debt finite amount. Volumes and prices of the new debt acquiring quickly accelerate, especially in the small neighborhood of the critical point. These considerations could be implemented as a new debt function being a solution to a Bernoulli differential equation. As it is well known, a simple Bernoulli process contains singularity. The latter represents the bubble burst which is inevitable result of herding. The bubble singularity can be explained alternatively via growing leverage in the market characterized by increasing asset prices. This phenomenon was thoroughly explained in (Adrian, Shin, 2008).

From the "pure financial" point of view, singularity could be explained as a natural consequence of interactions between debt and money. However large, but finite, amount of money becomes, at the critical point, negligibly small comparing to the infinitely large debt value. The same features are discernible in any actual crisis which is nothing more than the acute shortage of liquidity resulting from an eventual burst of a bubble. Like in the reality, to postpone the crisis at the second critical point, $s = s^*$, money issuance should be increased, even to the larger amounts than before the crisis. In this important aspect, as it seems, the model could explain paradoxical, at first glance, behavior of major central banks after the credit crunch 2007-09. instead of evaporating "excess liquidity" having been existed on the eve of the crisis, all the major central banks dramatically increased assets in their balance sheets, thus pumping trillions of dollars into the economy in several phases of "quantitative easing".

**Some conclusions**

The model shows, rather convincingly in our view, that unbounded credit expansion would lead to imminent debt bubbles and inevitable crises intertwined into financial cycles going on along the axis of money issuance. These cycles could be conceived as a natural consequence of an expansionary monetary policy performed by central banks in contemporary economies. One of the evident corollaries out of such a premise would be the following: in order to escape a systemic collapse it is necessary to change the monetary policy. The simplest thing to do it (in the model) would be by making the rate of money issuance equal to zero, $a = 0$. Pure random oscillations of money issuance, as it seems, would eliminate crises, thus curing the decease of debt cyclicity. So simple and so easy to prescribe the medicine in need, yet almost impossible to implement it!

The paradox is easily explained, though, for in the modern financial world such a medicine would be almost unanimously considered as an event being much worse than the decease itself. Anything, however remotely resembling some sort of a standard to be imposed upon the money issuance, would be rejected, almost surely. Contemporary history of money is a persistent process of credit expansion, invariably. Very lately, in spite of the fact that excess of money was the major factor of credit crunch in 2007-09, the post-crisis policy of the G4 central banks have been unambiguously an expansionary one (BIS,2012; Bloomberg, 2012). All the attempts of preventing the money and credit expansion seem to be totally unrealistic, undesirable and almost impossible. It is enough to mention, in this respect, either the just initiated the QE3 phase of the Fed policy, or the failed Bundesbank's opposion to the ECB latest plans of debt purchases. *Similia similibus curantur* (similar things are cured by the like). Understandably, all the deliberations of returning to the gold or other standards would be viewed, almost unanimously, as reckless speculations extremely harmful, under the circumstances, for the fragile buds of economic recovery. Even the simple 4% rule of monetary expansion (the number could be accommodated to the current economic conditions) which had been proposed by M. Friedman in the 70-ties, seems to be out of question. Hence, were the central banks to continue their unilateral defense of creditors, thus totally ignoring interests of borrowers, then, under the circumstances, the L. vonMises assertion of the crises inevitability would have to be very likely realized as a self-fulfilling prophecy.

**Смирнов А.Д.** Покупать или не покупать – вопрос не в этом: простая модель кредитной экспансии. Препринт WP7/2012/0… М: НИУ ВШЭ, 2012. – 42с.

В простой модели кредитной экспансии повышение ликвидности рынков меняет ориентацию инвесторов, заставляя их систематически завышать рыночные цены относительно истинной стоимости активов. Дивергенция цен формирует финансовый пузырь, динамика которого представлена стохастическими дифференциальными уравнениями денег и стоимости долга. Критические точки эмиссии ликвидности позволяют различать фазы кредитного цикла, конкретизируя гипотезу финансовой нестабильности Х. Минского, причём большим размерам эмиссии денег соответствуют большие вероятности дефолта системы. Перифраза известной дилеммы Гамлета раскрывает иллюзорность попыток инвесторов избежать краха в условиях растущей эмиссии денег, порождающей финансовые пузыри и кризисы. Без фатализма австрийской школы, модель поддерживает утверждение о том, что кредитная экспансия обязательно заканчивается системным кризисом, которому соответствует единичная вероятность дефолта.

**Смирнов Александр Дмитриевич** - заслуженный деятель науки РФ, доктор экономических наук, профессор, действительный член Российской академии естественных наук, Национальный исследовательский университет - Высшая школа экономики. 101987 Москва, Покровский бульвар, 11. Тел: 772 9590*2175.
E mail: adsmir@hse.ru , adsmir@gmail.com